\documentclass[12pt]{JHEP}

\usepackage{amsmath,epsfig}
\usepackage{amssymb,amsfonts}
\usepackage{latexsym}
\usepackage{graphicx}
\relax
\renewcommand{\theequation}{\arabic{section}.\arabic{equation}}

\def\hre#1#2{\href{http://arxiv.org/abs/#1/#2}{[ArXiv:#1/#2]}}

\def\be{\begin{equation}}
\def\ee{\end{equation}}
\def\bea{\begin{eqnarray}}
\def\eea{\end{eqnarray}}

\newcommand\fverb{\setbox\pippobox=\hbox\bgroup\verb}
\newcommand\fverbdo{\egroup\medskip\noindent%
                        \fbox{\unhbox\pippobox}\ }
\newcommand\fverbit{\egroup\item[\fbox{\unhbox\pippobox}]}

\newcommand{\la}{\lambda}

\newcommand{\bear}{\begin{eqnarray}}

\newcommand{\eear}{\end{eqnarray}}

\newbox\pippobox

\def\lab{\label}
\def\6{\partial}
\def\f{\Phi}
\def\a{\alpha}
\def\nn{\nonumber}

\def\half{\frac12}

\def\le{\left}
\def\ri{\right}
\def\cO{{\cal O}}

\def\C0{{\bf C_0}}
\def\Y0{{\bf Y_0}}
\def\G0{{\bf G_0}}

\def\m{\mu}
\def\n{\nu}

\def\sq
\def\a{\alpha}

\def\l{\lambda}
\def\g{\gamma}

\def\tr{{\rm Tr}}
\def\k{\chi}

\def\eps{\epsilon}

\def\la{\langle}
\def\ra{\rangle}
\def\o{\omega}
\def\vh{\vec{h}}
\def\bz{\begin{itemize}}
\def\ez{\end{itemize}}
\def\bn{\begin{enumerate}}
\def\en{\end{enumerate}}

\def\ka{\kappa}
\def\ch{{\rm ch}}
\def\sh{{\rm sh}}
\title{Continuous Hawking-Page transitions in Einstein-scalar gravity}
\author{Umut G{\"u}rsoy\\
\href{http://www1.phys.uu.nl/wwwitf}{Institute for Theoretical
Physics, Utrecht University; Leuvenlaan 4, 3584 CE Utrecht, The
Netherlands.}}

\date{\today}
\preprint{
SPIN-10/18\\
ITP-UU-10/20}      

\abstract{We investigate continuous Hawking-Page transitions in
Einstein's gravity coupled to a scalar field with an arbitrary
potential in the  weak gravity limit. We show that this is only possible in a singular limit
where the black-hole horizon marginally traps a curvature
singularity. Depending on the subleading terms in the potential, a
rich variety of {\em continuous} phase transitions arise. Our
examples include second and higher order, including the {\em
Berezinskii-Kosterlitz-Thouless} type. In the case when the scalar
is dilaton, the condition for continuous phase transitions lead to
(asymptotically) linear-dilaton background. We obtain the scaling laws of thermodynamic
functions, as well as the viscosity coefficients near the
transition. In the limit of weak gravitational interactions, the
bulk viscosity asymptotes to a universal constant, independent of
the details of the scalar potential. As a byproduct of our analysis we obtain a one-parameter 
family of kink solutions in arbitrary dimension $d$ that interpolate between AdS near the boundary and 
linear-dilaton background in the deep interior.
The continuous Hawking-Page transitions found here serve as holographic models for  
normal-to superfluid transitions.}


\begin{document}

\maketitle

\section{Introduction}

Thermodynamics of radiation in asymptotically Anti-de-Sitter
spaces is quite interesting. It was shown by Hawking and Page in
\cite{HP} that, as one heats up a gas of gravitons in the global
AdS space-time, one finds a phase transition into nucleation of
black-holes at a finite temperature $T_c\sim \ell^{-1}$ where
$\ell$ is the AdS radius. At low temperatures, the only
equilibrium solution to Einstein's equations with asymptotically
AdS boundary condition is thermal radiation. Above a certain
temperature $T_{min} \sim \ell^{-1}$ a ``small black-hole" and a
``big black-hole" saddle coexist with the thermal gas (TG). Right
above $T_{min}$ these solutions have free energy larger than the
radiation, therefore they eventually tunnel into thermal
radiation. However, above a certain value $T_c> T_{min}$ the big
black-hole solution becomes the global minimum and the thermal gas instead
becomes quantum mechanically unstable against tunneling into the
big black-hole.\footnote{The small BH solution has negative
specific heat and it is always subdominant with respect to the big
BH and TG saddles. It will play no role in what follows.} This is
the so-called Hawking-Page transition in asymptotically AdS
space-times.

Witten in \cite{Witten} proposed that the Hawking-Page transition
in the $AdS_5\times S^5$ solution of IIB string theory corresponds
to a confinement-deconfinement type transition in ${\cal N}=4$
Yang-Mills theory on $S^3$ with radius $R\sim \ell$, through the
AdS/CFT corespondence\cite{malda}. There is a transition at
some $T_c\sim R^{-1} \sim \ell^{-1}$ and it only occurs in the
large color limit $N\to\infty$, where the number of degrees of
freedom becomes infinite. A more interesting case is when the
field theory is defined on $R^3$, which can be obtained as the
large radius limit of the sphere $R\to \infty$. Clearly $T_c \to
0^+$ in this case. The corresponding statement in the dual field
theory is that the ${\cal N}=4$ sYM theory is conformal, and when
placed on $S^1\times S^3$ with $S^1$ being the time-circle with
radius $1/T$, the thermodynamic quantities can only depend on the
dimensionless ratio $TR$. Therefore, there cannot be any
interesting thermodynamics in the flat space limit.

The situation changes, when there is an additional intrinsic scale
in the problem that does happen in non-conformal theories. A
basic example is  pure Yang-Mills theory with gauge group
$SU(N)$. The theory has running coupling, hence a dynamical scale
$\Lambda$ is generated through dimensional transmutation. Indeed,
lattice studies provide strong evidence for a second order
deconfinement transition for $N=2$ and a first order deconfinement
transition for $N>2$ in 3 space dimensions roughly at $T_c\sim
\Lambda$.

By a natural generalization of the ideas in \cite{Witten}, one
expects that Hawking-Page type transitions in asymptotically AdS
space-times dual to Yang-Mills type non-conformal
theories\footnote{By a Yang-Mills type theory we mean a
gauge-theory that exhibits quark confinement in the ground state.}
should correspond to confinement-deconfinement transitions in
these theories. In \cite{GKMN1}, it was shown that the simple
set-up of Einstein gravity coupled to a scalar field $\f$ with
some potential $V(\f)$ allows for asymptotically AdS type
solutions with a non-trivial intrinsic scale $\Lambda$ that is
generated due to the running scalar field. It was proposed in
\cite{GK}\cite{GKN} that the ground state (zero T) solution
generates color confinement in the dual field theory, if and only
if the IR asymptotics of the potential obeys, \be\lab{Vconf} V(\f)
\to e^{\frac43 \f} \f^P, \quad P\ge 0, \qquad as\,\,\,
\f\to\infty, \ee in 5 dimensions. A background that solves a
potential of the form (\ref{Vconf}) in the IR and becomes
asymptotically AdS in the UV indeed exhibits a Hawking-Page
transition that corresponds to the deconfinement transition in the
dual gauge theory.

In fact, the thermodynamics in the generic case $P>0$ is very
similar to the global AdS case that we described above: There is
always a thermal gas solution saddle at every T. This is the only
equilibrium configuration up to a certain $T_{min}$. Above
$T_{min}$ there are two black-hole solutions in addition to the
thermal gas: the small and the big BHs, just as in global AdS. At
a certain $T_c>T_{min}$ the big BH solution becomes the global
minimum of the theory. This temperature corresponds to the
deconfinement temperature in the proposed dual YM theory, where
the glueballs deconfine their glue contents. For $P>0$, one can
quite generally show \cite{GKMN2} that, there is a finite latent
heat and the transition is of first order.

In this paper, we ask the general question, ``under what
conditions gravity exhibits continuous Hawking-Page type phase
transitions?" By the word {\em continuous}, we mean second or
higher order transitions where the entropy is continuous across
the transition. In other words, the latent heat vanishes.
Regardless of the matter content of the gravity theory, the basic
condition for a continuous HP transition can easily be obtained in
the classical limit $G_D\to 0$. In this limit we arrive at the
(more or less obvious) fact that, {\em a continuous transition in
this limit can only occur in a singular region of space-time,
where the BH horizon marginally traps a curvature singularity}.
This is easily understood: in the weak curvature limit, the
contribution of thermal radiation to the entropy is negligible,
thus the entropy of the TG solution is very small (zero in the
strict $G_D\to 0$ limit). Then, continuity of entropy requires
that the horizon size of the BH solution becomes very small as one
approaches the transition temperature from above.

To be more definite, we approach the problem of continuous
transitions in the context of Einstein's gravity coupled to a
scalar. In fact,  it was already observed in \cite{GKMN1} that,
the aforementioned HP transition may be continuous in the {\em
marginal limit} $P=0$ in (\ref{Vconf}). Here, we focus in this
particular case in more detail and also generalize the analysis to
arbitrary $d+1$ dimensions. Our general result can be stated as
follows: there exists a continuous Hawking-Page transition in
$d+1$ dimensional Einstein-scalar theory if and only if
\be\lab{Vhp} V(\f) \to e^{\frac{4}{d-1} \f} \le( 1+
V_{sub}(\f)\ri), \qquad as\,\,\, \f\to\infty, \ee where $V_{sub}$
denotes subleading corrections that vanish in the deep IR. We also
find that {\em these corrections determine the type of the
transition}: For an exponential fall-off in $V_{sub}$, one finds
nth order transitions for $n\ge 2$. In this case the free energy
difference scale as, \be\lab{sca1} \Delta F \sim (T-T_c)^n, \qquad
as\,\,\, T\to T_c^+.\ee On the other hand, for a power-law
fall-off, a Berezinskii-Kosterlitz-Thouless type scaling arises:
\be\lab{sca2} \Delta F \sim e^{-A t^{-\frac{1}{\a}}}, \qquad
as\,\,\, T\to T_c^+,\ee where $\a>1$ is a constant.

A good indication that the marginal case $P=0$, corresponds to
continuous transitions is obtained by investigating the
equilibrium solutions of $V_{conf}$. As mentioned above, for $P>0$
generically there are three solutions at high temperatures: the
thermal gas, the big and the small black-holes. In the marginal
limit $P\to 0$ the small black-hole solution disappears.
Therefore, in the high T regime there exist only the TG and the
big BH branches. In fact, in this case $T_{min}$ and $T_c$ can be
shown to coincide \cite{GKMN2} and the thermodynamics becomes much
simpler than the generic $P>0$ case: The low T (high T) phase is
dominated by the TG (big BH) saddle and these phases are separated
by a Hawking-Page transition at $T_c$. This is precisely what one
expects in a continuous phase transition: At $T_c$ the new saddle
is created and immediately becomes the global minimum of the free
energy. Indeed the small BH solution can be considered as a kink
that connects the TG and the big BH solutions which can only exist
in first order transitions. In the limit where the transition
becomes continuous this kink should disappear.

As we mentioned above, in the classical limit, the continuous
transitions generically occur in a singular region where the
horizon area becomes very small. In the particular example of the
Einstein-scalar gravity with potential (\ref{Vconf}) and with
$P=0$, this region corresponds to $\f_h\to \infty$ where $\f_h$ is
the value of the scalar on the horizon\footnote{It can also be
viewed as the horizon location in a coordinate system where the
radial direction is given by $\f$.}. In this limit, it was shown
in \cite{GKMN2} that the BH and the TG solutions become
degenerate, that is indeed another basic property of second or
higher order transitions. In fact, in the strict classical limit
$G_D\to 0$, the transition point corresponds to $\f_c=\infty$.

Clearly we cannot trust Einstein's gravity in this singular
region. On the other hand we would like to study the scaling of
thermodynamic functions with $T-T_c$, an information contained in
this singular region. We avoid this problem here by fine-tuning
the sub-leading term in (\ref{Vhp}) such that the region
$T\lesssim T_c$ corresponds to a wide range $\f_h\in (\f_s,
\infty)$ where $\f_s$ is the value of $\f$ where the scaling
behavior in (\ref{sca1}) or (\ref{sca2}) sets in. This is clearly
a crude method that we can resort to, only in a phenomenological
approach where we use the freedom in the choice of the potential
to hide the problem of singularity. This method serves our
purposes in this paper, that is to work out the qualitative
behavior of thermodynamic functions in a continuous HP transition.
However, a more detailed future study should take into account the
$\a'$ corrections by investigating continuous HP transitions in
the full string theory.

Studying the IR geometry of the solutions to (\ref{Vhp}) we arrive
at another interesting result: the BH and the TG geometries both
asymptote to a linear-dilaton geometry, in the case when the
scalar field is the dilaton. More precisely, the ground state (TG
solution) that uniquely follows from the potential (\ref{Vhp}) is
asymptotically linear-dilaton in the deep interior of the
geometry. At the same time, the BH geometry approaches to the TG
geometry in the limit $T\to T_c$. Therefore both of the saddles
are of  linear-dilaton type in this region of space. It is
well-known that the linear-dilaton geometry is an $\a'$ exact
solution to the full string theory. This observation may prove
important in studying the analogous problem in the full string
theory.

In the limit where we can ignore gravitational interactions and
higher derivative corrections, Einstein's gravity predicts certain
{\em universal results} around a continuous transition. One such
quantity is the value of the bulk viscosity per entropy as $T\to
T_c$. We find that this value is constant across the transition
\footnote{See \cite{Buchel} for another example where $\zeta/s$ is
found to be constant.}: \be\lab{zetapers} \frac{\zeta}{s}\to
\frac{1}{2\pi(d-1)}, \qquad as\,\,\, T\to T_c. \ee The result is
universal in  the sense that it is independent of the details of
the theory such as the choice for the scalar potential $V(\f)$, the transition
temperature $T_c$ etc., and it is only subject to $\a'$ corrections
in a gravity theory with higher derivative terms.

We should mention that although this work can be regarded
as an investigation purely  in gravity, we have holographic
applications in mind. In this paper, we shall not specify the
expected dual field theory. In fact most of our findings will not
depend on the UV geometry either. Yet, we would like to have an
asymptotically AdS geometry in the UV, to be able use the
prescriptions of the AdS/CFT duality in holographic applications.
A particularly interesting such application is studied in a
companion paper \cite{exotic2}, where we propose that the
continuous phase transitions found here are holographically
related to normal-to-superfluid transitions in spin-models.

In the course of searching for backgrounds with AdS asymptotics, we also obtain a one-parameter 
family of analytic kink solutions that flow from an AdS extremum to the linear-dilaton 
background in the string frame. 

In the next section, we list the general conditions for the
Einstein-scalar theory to exhibit a continuous Hawking-Page
transition. In section 3, we present the classification of the 
transitions that follow from different fall-off behavior
of $V_{sub}$ in (\ref{Vhp}). In this section, we also present a
coordinate invariant method to study thermodynamics in
Einstein-scalar gravity, that was first constructed in
\cite{GKMN2}. Here we generalize this method here to arbitrary
dimensions. Section 4 is devoted to derivation of the scaling
behavior of thermodynamic functions: the free energy, the entropy,
the specific heat and the speed of sound. In section 5 we study
the shear and bulk deformations around the black-hole solution and
study the viscosity-entropy ratios $\eta/s$ and $\zeta/s$ in the
transition region. In section 6 we construct the aforementioned analytic kink soutions. 
Finally in section 7, we briefly mention what type of
corrections are expected in our findings when the set-up is
embedded in full string theory. Finally we present a discussion on
our results and an outlook for future directions in the last
section. Six appendices detail our calculations.

\section{Weak phase transitions and small black-holes}
\lab{section2}

\subsection{Generalities}
\lab{gens}

In this section we determine the requirements for presence of continuous phase transitions
in the Einstein-scalar theory in $d+1$ dimensions, defined by the action:
\begin{equation}\label{action}
 {\cal A} = \frac{1}{16\pi G_D} \int d^{d+1}x \sqrt{-g}\le( R - \xi (\6\f)^2 + V(\f)
  \ri)+\,\,\, G.H.
\end{equation}
We keep the normalization of the scalar kinetic term $\xi>0$ as a
free parameter for later convenience. The last term, that we shall not need to specify here, is the
Gibbons-Hawking term on the boundary. $G_D$ is the Newton's constant.
The total number of bulk dimensions is denoted by $D=d+1$.

There are only  two types of backgrounds at finite T (Euclidean time is
compactified), with  Poincar\'e symmetries in $d-1$ spatial dimensions, and and additional $U(1)$ symmetry in the  Euclidean time direction:
the {\it thermal graviton gas},
\begin{equation}\label{TG}
  ds^2 = e^{2A_0(r)}\le( dr^2 + dx_{d-1}^2 + dt^2\ri), \qquad \f= \f_0(r),
\end{equation}
and the {\it black-hole},
\begin{equation}\label{BH}
  ds^2 = e^{2A(r)}\le( f^{-1}(r)dr^2 + dx_{d-1}^2 + dt^2 f(r) \ri), \qquad \f=
  \f(r).
\end{equation}
We define the coordinate system such that the boundary is located at $r=0$.
For the potentials $V$ that we consider in this paper, there is a curvature singularity in the deep interior,
at $r=r_s$.  In (\ref{TG}), r runs up to singularity $r_s$. In
(\ref{BH}) there is a horizon that cloaks this singularity at $r_h<r_s$ where $f(r_h)=0$.
t is the Euclidean time that is identified as $t\sim t+1/T$. This defines the temperature T of the
associated thermodynamics. In the black-hole solution, the relation between the temperature and $r_h$ is obtained in the standard way, by assuming that the near horizon geometry is conical and the conical singularity is absent:
\begin{equation}\label{Trh}
4\pi T  = -f'(r_h).
\end{equation}
This identifies T and the surface gravity in the BH solution.

The geometries (\ref{TG}) and (\ref{BH})  are the only backgrounds
with  the boundary condition that $A(r)\to A_0(r)$, $\f(r) \to
\f_0(r)$
 and $f(r)\to 1$ near the boundary\footnote{This is somewhat rough,
 see \cite{GKMN2} for precise matching conditions in the variable
$r$, near the boundary. The easiest way to implement the matching
is to consider a coordinate frame where $\f$ is the radial
variable and identifies $\f = \f_0$ in the entire range of the two
solutions. Then the only nontrivial condition is $A(\f) \to
A_0(\f)$ and $f(\f) \to 1$ near the boundary. This is what we will
do in section \ref{class}.}. We are interested in determining the
necessary and sufficient conditions for existence of a {\em
continuous} Hawking-Page type transition between (\ref{TG}) and
(\ref{BH}). We define a {\em continuous phase transition} as the
one where the entropy difference vanishes $\Delta S(T_c) = 0$ at
the transition. This defines phase transitions of second and
higher order, including infinite order, such as the BKT transition
\cite{BKT}\footnote{See \cite{Son} and \cite{Liu} for recent
papers where the BKT scaling was observed in the context of
condensed matter applications of holography. These papers consider
quantum phase transitions however, where the scaling is not in T
but in a quantum coupling.}.

Thermodynamics is determined by the Gibbs free energy that is given by the on-shell value of the action (\ref{action}).
Taking the difference between the on-hell values on the BH and the TG solutions 
removes possible divergences due to infinite volume of the geometries and yields the regularized action:
\begin{equation}\label{GS}
T^{-1}\Delta G(T) =  {\cal A}(BH) - {\cal A}(TG) + \cdots
\end{equation}
where the ellipsis refer to corrections that are $\cO(G_D)$ suppressed with respect to the first term.

The Newton's constant in (\ref{action}) is important in determining the overall size of the thermodynamic functions.
It also plays an important role in the correspondence between gravity and a dual field
theory, where $G_D$ should be proportional to the number of
degrees of freedom in the dual quantum field theory. This is most
clear when the proposed dual theory is is a gauge theory with $N$
colors:
\begin{equation}\label{Mp}
\frac{1}{16\pi G_D} = M_p^{d-1} N^2,
\end{equation}
where $M_p$ is a ``normalized" Planck scale, that is generally of
the same order as the typical curvature of the
background\footnote{For example, in the improved holographic QCD
models \cite{GKMN1}, it is given by $M_p = (45\pi^2)^{-\frac13}
\ell^{-1}$ where $\ell$ is the curvature radius of the
asymptotically AdS geometry near the boundary.}. On the gravity side the factor N arises in a similar fashion 
to the $AdS_5 \times S^5$ case. Here it is supposed to be proportional to the RR $d+1$ form flux on the geometry \cite{GK}. 
As explained in \cite{GK} the space-filling flux can be integrated out by replacing it with its equation of motion
in the action, and it only contributes to the potential  $V(\f)$. The potential $V$ that is written in (\ref{action}) is supposed 
to include this contribution. This procedure is completely analogous to the 5-dimensional case discussed in \cite{GK} 
and we refer the reader to this paper for details.  

Clearly, in order
to be able to ignore interactions between bulk fields one needs to
take the large N limit. This can easily be done when the scalar is dilaton: in an asymptotically AdS solution where the boundary value of the 
dilaton is constant $\f_0$, the large N limit is given by sending the dilaton  $\f_0\to -\infty$, and the flux $N\to\infty$ such that $e^{\f_0} N = const.$


In this work, we will mostly be interested in  the scaling of the various thermodynamic
quantities with T and {\em not} in the overall size of these quantities.  For that purpose, it is more convenient to factor out the overall $G_D$ dependence by dividing the free energy by $V_{d-1} N^2$ where $V_{d-1}$ is the volume of the spatial part, and $N^2$ is defined by (\ref{Mp}). We, thus define the  normalized free energy {\em density},
\begin{equation}\label{FG}
\Delta F(T) =  \frac{1}{V_{d-1} T^{-1} N^2}\le({\cal A}(BH) -
{\cal A}(TG)\ri) + \cdots \equiv F_0(T) + \delta F(T).
\end{equation}
Here $F_0$ is defined as the leading $\cO(N^0)$ piece in the free energy density, and $\delta F$ refers to the $\cO(N^{-2})$ corrections.
The large limit $N\to\infty$ ($G_D\to 0$)  defines
a saddle point approximation from the gravitational point of view. The corrections that we denote by $\delta F$ in (\ref{FG}) arise, for finite but small $G_D$ from the
determinant of gravitational fluctuations around the saddle:
\begin{equation}\label{logdet}
\delta F(T) = \tr\log O_{BH} - \tr\log O_{TG},
\end{equation}
where $O$ is the d'Alembertian operator of the fluctuations
around the saddle geometries.\footnote{In principle, for finite $G_D$, one has
to make sure that this difference is free of divergences. By the
boundary condition that BH asymptotes to TG near the boundary, the
two spectra in the UV are guaranteed to be the same (this can be
checked by a WKB analysis), hence the difference should be 
convergent.}

Although, we shall mostly be interested in the leading $F_0$ term in (\ref{FG}), we note the following
important consequence of the $\delta F$ correction: Using the 1st law, the entropy (density)
difference is given by, \footnote{The backgrounds that we consider
in this paper satisfy the required energy conditions and the laws
of thermodynamics should apply.}
\begin{equation}\label{ent1}
\Delta S = S_{BH} - S_{TG} = - \frac{dF_0}{dT} - \frac{d\delta F}{dT},
\end{equation}
where we normalize the entropy the same way as in (\ref{FG}). The
$S_{TG}$ term above has no leading $\cO(N^0)$ piece because the TG
geometry has no horizon. Therefore, one has,
\begin{equation}\label{ent2}
 \Delta S = S_{BH} - \frac{d\delta F}{dT}.
\end{equation}
We shall ignore the $1/N^2$  corrections throughout the paper and drop the second term above, hence the entropy (density) difference of the dual theory is entirely determined by the black-hole entropy
that is therefore given by the horizon size at $r_h$:
\begin{equation}\label{ent3}
\Delta S = S_{BH} = \frac{e^{(d-1)A(r_h)}}{4G_D N^2}  = 4\pi M_p^{d-1} e^{(d-1)A(r_h)},
\end{equation}
where we used the relation (\ref{Mp}) in the second equation.

Therefore, we learn from (\ref{ent3}) that---up to $1/N^2$ corrections that we neglect in this paper---{\em a necessary condition for a continuous phase transition is that
the black-hole becomes infinitesimal at $T_c$.} This is an {\em infinitesimal black-hole}\footnote{Not to be confused with the aforementioned ``small black-hole'' saddle which is a black-hole with a negative specific heat. In all of the examples we consider in this paper,
we only consider black-holes with non-negative specific heat.} with vanishing  {\em classical} horizon size.\footnote{Note that the black-hole at $T_c$ should be non-extremal, but with vanishing enthalpy.}
In the backgrounds we consider there is a single curvature singularity at $r=r_c$. Precisely at $T_c$, the singularity is marginally trapped by the black-hole, and
$T_c$ corresponds to the temperature where the horizon approaches the singularity $r_h\to r_c$ \footnote{One may consider the possibility of string corrections at this point.
We will discuss such possible corrections in the last section.}.

There are further conditions for the existence of a phase transition as explained in the following.
Before we present the  derivation in the next subsection, let us state the final result:
{\em In the Einstein-scalar gravity, there exists a continuous phase transition into nucleation of black-holes if and only if
the scalar potential $V(\f)$ has the asymptotic behavior,}
\be\lab{vsum}
V(\f) \to e^{2\sqrt{\frac{\xi}{d-1}}\f}, \qquad \f\to\infty.
\ee
This result is of course subject to field redefinitions in the action (\ref{action}). Below, we present a derivation where we stick to the form of the action given in (\ref{action}).

\subsection{Conditions for presence of continuous transitions}
\lab{condts}

We are interested in a Hawking-Page type phase transition when the
thermal gas of gravitons (\ref{TG}) becomes unstable against
nucleation of black holes (\ref{BH}) and we ask the question {\em
under what conditions this transition is higher than first order?}
The thermodynamic requirements are clear:
\begin{itemize}
  \item[i.] $\exists\,\,\, T_c$ such that $0<T_c<\infty$ and
  $\Delta F(T_c)=0$,
  \item[ii.] $\Delta S(T_c)=0$.
\end{itemize}
Solution of these conditions for a generic scalar potential $V$
in (\ref{action}) may seem rather arbitrary. This is not the case
however. As we show below, the conditions above imply the specific
form (\ref{vsum}) for the asymptotic form of $V$.

Before solving the conditions above, first we have to define the
various thermodynamic functions in terms of geometric quantities.
We solve the equations of motion and derive the thermodynamics 
in a coordinate independent approach in section
\ref{XYvar}. Here we shall first work with the coordinate frame---that we
call the r-frame---defined by the metrics (\ref{TG}) and
(\ref{BH}) for simplicity. In the r-frame, one derives the
following Einstein and scalar equations of motion from
(\ref{action}):
\bea \label{E1}
  A'' - A'^2 + \frac{\xi}{d-1} \f^{'2} &=& 0, \\
\lab{E3}
f'' + (d-1)A'f' &=& 0, \\
\label{E2}
   (d-1)A'^2f + A'f'+A''f -\frac{V}{d-1} e^{2A} &=& 0 .
\eea
One easily solves (\ref{E3}) to obtain the ``blackness function"
$f(r)$ in terms of the scale factor as,
\begin{equation}\label{f}
  f(r) = 1- \frac{\int_0^r e^{-(d-1)A}}{\int_0^{r_h} e^{-(d-1)A}}.
\end{equation}
Then the temperature of the BH is given by equation  (\ref{Trh}):
\begin{equation}\label{T}
  T^{-1} = 4\pi e^{(d-1)A(r_h)}\int_0^{r_h} e^{-(d-1)A(r)}dr.
\end{equation}
The difference between the entropy densities of the BH and the TG
solutions is given by the BH entropy density (\ref{ent3}) up to
$1/N^2$ corrections that we ignore from now on:
\begin{equation}\label{S}
\Delta S = \frac{1}{4G_D N^2}e^{(d-1)A(r_h)}.
\end{equation}
 The difference in the free energy densities (\ref{FG})
 can equally well be evaluated by integrating  the first law of thermodynamics
 (\ref{ent1}), \cite{GKMN2}:\footnote{We ignore the $\delta F$ term in (\ref{ent1}) that
 is subleading in $N$.}
\begin{equation}\label{F}
  \Delta F(r_h) = -\frac{1}{4G_D N^2} \int_{r_c}^{r_h}
  e^{(d-1)A(\tilde{r}_h)}\frac{dT}{d\tilde{r}_h} d\tilde{r}_h.
\end{equation}
Here $r_c$ is the value of the horizon size that corresponds to
the phase transition, $T(r_c) = T_c$, where the difference in free
energies should vanish. We note that, dependence on $r_h$ in the
integrand of (\ref{F}) is two-fold: There is the explicit
dependence on the variable $r_h$ and there is the implicit
dependence that arise because the function $A(r)$ itself changes
as $r_h$ is varied\footnote{This is clear when solving (\ref{E1})
by specifying the boundary conditions at $r_h$. See section
{\ref{XYvar}} for more detail.}. As we showed in \cite{GKMN2},
this implicit dependence on $r_h$ is suppressed in the limit $r_h$
close to the singularity $r_s$. This is a fact that will prove
important in the analysis of the next section.

It was demonstrated in \cite{GKN} that, for any domain-wall type
geometry of the form (\ref{TG}) with the condition that the scale factor obeys $A'(r_0)<0$ near the
boundary\footnote{This is the only technical requirement in this
section. We note that asymptotically AdS geometries are of this
type.}, the possible asymtotics in the deep interior can be
classified as:
\begin{enumerate}
\item[A.] An asymptotically AdS geometry in the deep interior as $r\to
\infty$, \item[B.] A curvature singularity at finite $r= r_s$
where $\phi_0\to \infty$ and $A_0\to -\infty$, and
\item[C.] A curvature singularity at $r= \infty$ where
$\phi_0\to \infty$ and $A_0\to -\infty$.
\end{enumerate}
We want to determine which type of geometry above admits a
Hawking-Page transition by satisfying the conditions i and ii stated in
the beginning of this section.

We first focus on the condition ii: For a {\em continuous} phase
transition, one should require that the entropy difference
(\ref{S}) should vanish at the transition point $r_c$. Then, from
the classification above we learn that $r_c$ should correspond to
either of the {\em singular points} $r=r_s$ in case ii, or
$r=\infty$ in the cases i or iii, because that is the only point
where the scale factor vanishes, so does $S$ by
(\ref{S}).\footnote{We refer to \cite{GKMN2} for a derivation of
the fact that the BH(TG) metric functions $A$($A_0$) become
asymptotically the same at the singular points. Thus requirement
that $\exp{A_0}$ vanishes is equivalent to the requirement that
$\exp{A}$ vanishes.}

Now we consider the condition ii above. In order to maintain a
{\em finite} $T_c$, from (\ref{T}) one should require that the
metric function near the singularity should behave linearly:
\begin{equation}\label{As}
  A(r) \to -A_\infty~r + \cdots, \qquad r\to\infty
\end{equation}
where $A_\infty$ is a constant and the ellipsis may involve
(powers of) log's or terms that are sub-dominant as $r\to\infty$.
This singles out the deep-interior geometry of the type C above.
Then, the rest of the condition i is satisfied automatically: The
fact that $\Delta F$ vanishes at $r_c=\infty$ follows immediately
from the definition $F(T_c) = E(T_c) - T_c S(T_c)$. The enthalpy
piece vanish because the entropy vanishes and $T_c$ is finite. The
energy also vanish by the fact that in the limit $T\to T_c$ which
corresponds to $r_c=\infty$ the BH becomes of zero size, hence the
ADM mass vanishes\footnote{See \cite{GKMN2} for a careful
computation of the ADM mass.}.

This is of course in accord with the first law of thermodynamics
(\ref{F}) where we tacitly made the assumption that $\Delta F$
vanishes at $r_c$. Here, in addition to this we learned that, {\em
for a continuous transition}, one needs $r_c = \infty$ and one should
require (\ref{As}) in the deep-interior. This latter condition can
be expressed in a coordinate-invariant way, as a condition on the
asymptotic form of the scalar potential $V(\f)$, as follows: The
Einstein equation (\ref{E1}) determines the asymptotic form of the
scalar field:
\begin{equation}\label{fs}
  \f(r) \to \sqrt{\frac{d-1}{\xi}} A_\infty~r + \cdots, \qquad r\to\infty
\end{equation}
where, as in (\ref{As}) the ellipsis may involve logs. Now, the
asymptotic form of the potential $V$ is determined by the scalar
e.o.m., (\ref{E2}), using (\ref{As}), and inverting (\ref{fs}) in
favor of $\f$ as,
\begin{equation}\label{Vs}
  V(\f) \to V_\infty~e^{2\sqrt{\frac{\xi}{d-1}}\f}\le(1 + V_{sub}(\f)\ri), \qquad \f\to\infty
\end{equation}
where $V_\infty$ is a constant.

The term $V_{sub}$ in (\ref{Vs}) is a sub-leading piece:
$V_{sub}\to 0$ as $\f\to\infty$. {\em This piece will be important
in determining the order of the phase transition in the next
section.} It is interesting to note that, even if log terms were
present in (\ref{As}) and (\ref{fs}), they conspire to produce
exactly the single exponential (with no $\log\f$ corrections in
the exponential) in the leading asymptotics.

The only dimensionful parameter in the problem is $V_\infty$ in
(\ref{Vs}). Indeed, $V$ has dimensions of $\ell^{-2}$ where $\ell$
is a typical radius of curvature of the TG geometry, for example ,
it is the AdS radius when the TG geometry is asymptotically AdS.
Thus, all of the other dimensionful quantities above, i.e.
$A_\infty$, $\f_\infty$ and $T_c$ should be determined in terms of
$V_\infty$:
\begin{equation}\label{relsinf}
  T_c \propto A_\infty \propto \f_\infty \propto \ell^{-1} \propto
  \sqrt{V_\infty}.
\end{equation}
We shall determine the proportionality constants precisely in the next section.

{\bf To summarize:} In the Einstein-scalar set-up, continuous
phase transitions of order $\ge$ 2 arise with and only with the
asymptotics (\ref{Vs}). The transition happens at $r_h=r_c=\infty$
where the entropy, the energy and the free energy vanishes, as the
black hole has vanishing {\em classical} horizon in this
limit\footnote{As explained in section \ref{gens} above, vanishing
of the BH entropy and free energy density is equivalent to the
vanishing of the entropy and free energy {\em difference} between
the BH and the TG backgrounds, to leading order in $1/N^2$.}.
However, the surface gravity, hence the temperature remains
finite. Finally, the value of the transition temperature is
controlled by constant $V_\infty$ in (\ref{relsinf}).

{\bf A Comment:} We have a BH geometry that has vanishing mass at
$T_c$, to leading order in the $1/N^2$ expansion. Yet the
temperature $T_c$ is finite. One should ask how a massless object
can have finite temperature? Does it radiate? In fact our
situation is very similar to what happens in the massless limit of
linear dilaton black-holes \cite{ldBH}. As the authors of
\cite{ldBH} show in the special case of a massless linear dilaton
BH, although the temperature remains finite, the radiation from
the event horizon vanishes in the massless limit, hence there is
no contradiction. In our case as well, one can show that the
radiation vanishes in the massless limit $r_h\to\infty$. We will
show this in section \ref{viscdisp} by computing the shear and the
bulk viscosity coefficients $\eta$ and $\zeta$ and showing that
they vanish in the limit $r_h\to\infty$. However, we also show
that the non-trivial and interesting quantities are
not the {\em bare} viscosity coefficients $\eta$ and $\zeta$
but the normalized ones $\eta/S$ and $\zeta/S$ which remain finite
in the limit $T\to T_c$. We refer the reader to section
\ref{viscdisp} for more details.

\subsection{Linear dilaton background in NCST}
\lab{lindil}

Now, we consider the special case when the scalar $\f$ is the
dilaton of $(d+1)$ non-critical string theory. The NS-NS action of
NCST in $d+1$ dimensions read,
\begin{equation}\label{saction}
  S \propto \int d^{d+1}x \sqrt{-g_s} e^{-2\f} \le( R_s + 4
  (\6\f)^2 + \frac{\delta c}{\ell_s^2}+\cdots \ri),
\end{equation}
where we only show the relevant terms in the two-derivative sector, the ellipsis refer
to higher order terms in the derivative expansion. $\delta c$ refers to the deficit central charge
in $d+1$ dimensions, e.g. it is $9-d$ in a parent 10D critical string theory.
The NS-NS two-form B is chosen to constant, hence it does not contribute to the equations of motion. 
The subscript s refers to string frame
quantities. 

One passes to  the Einstein frame by the Weyl
transformation $g_s^{\m\n} = e^{4\f/(d-1)}g^{\m\n}$. The
transformation produces an Einstein-frame action of the form
(\ref{action}) with the normalization $\xi$ fixed as,
\begin{equation}\label{xincst}
  \xi_s = \frac{4}{d-1}.
\end{equation}
In this case the requirement for weak phase transitions (\ref{Vs})
become,
\begin{equation}\label{Vsncst}
  V\to V_\infty~e^{\frac{4}{d-1}\f}+\cdots
\end{equation}
Interestingly, the form of the dilaton potential (in the Einstein
frame), (\ref{Vs}) that follows from the condition for existence
of continuous phase transitions, precisely coincides with the
dilaton potential that one obtains in the NCST after passing to
the Einstein frame. At the same time, we know that there is an
$\a'$-exact solution to the NCST, that is {\em the linear dilaton
background.} \cite{PolchinskiBook}. Indeed the asymptotic form of
the solutions we found in the previous section, both for the TG
(\ref{TG}) and the BH (\ref{BH}) backgrounds\footnote{In the BH
case the blackness function $f(r)$ approaches 1 as the horizon
size becomes smaller \cite{GKMN2}.}, become linear dilaton
geometries in the limit $r_h\to \infty$, if one passes to the
string frame:
\begin{equation}\label{ld}
  ds_s^2 = \le(dr^2 + dx^2_d \ri) \le(1+\cdots\ri),\qquad  \f(r)
  = \frac{d-1}{2} A_\infty~r (1 + \cdots)
\end{equation}
where the corrections denoted by the ellipsis vanish as
$r\to\infty$. The precise form of these corrections depend on
$V_{sub}$ in (\ref{Vs}).

The crucial fact is that when $\f$ is the dilaton of NCST, the
string-frame scale factor $A_s = A + 2\f/(d-1)$ vanishes when $V$
has the form (\ref{Vsncst}), thus one obtains a linear-dilaton
geometry in the deep-interior, in the string frame. Because the linear-dilaton is an exact solution to non-critical string theory in all orders in $\a'$
we expect that the leading term in the condition (\ref{Vs}) for presence of
continuous transitions is $\a'$-exact. We have more to say on
this in section \ref{stcor}.

In passing, we note that in this particular case of the D-dim.
NCST (with $\xi=4/d-1$), the coefficients $A_\infty$, $\f_\infty$
in (\ref{As}) and (\ref{fs}), and the transition temperature $T_c$
can be expressed in terms of the parameter $V_\infty$ in a precise
way (see App. \ref{vinftc} for derivation):
\be\lab{inftc} A_\infty = \frac{2}{d-1}\f_\infty =
\frac{4\pi}{d-1}~T_c = \frac{\sqrt{V_\infty}}{d-1}. \ee
This last equation, in particular, grants us with direct control
over the transition temperature by adjusting the amplitude of the
scalar potential in the asymptotic region. Clearly, this is very
much desired in phenomenological applications.

\subsection{The problem of singularity}
\lab{singular}

We showed above that existence of a continuous Hawking-Page
transition in the Einstein-scalar theory requires a singular
geometry with a curvature singularity in the deep interior. The
transition happens at the point when the horizon {\em marginally
traps the singularity.} 

Of course, the Einstein's theory breaks down as one approaches
this singular limit and one can no longer control the
calculations. In fact, in the string frame we do not have to worry about the curvature corrections
precisely around the transition region because the curvature invariants can be shown to 
vanish in this limit, see Appendix \ref{AppRicci}. Instead the issue is that the dilaton becomes
very large hence the effective string coupling becomes strong in this region and our assumption of 
ignoring the gravitational interactions cannot be justified. We can avoid this problem, if 
we have the means of fine-tuning the  asymptotic value of $e^{\f}$ near the boundary to be parametrically small. 
This is what we shall do in the following: we will adjust the
dilaton potential $V$ in such a way that the approach to the
singularity is slow enough. More precisely, we will choose the
subleading terms in the potential such that the vicinity around
$T_c$ corresponds is away from the large $\f$ region.
How to manage this, is explained in
section \ref{finetune}. In section 6 
we construct explicit examples where we can show that this fine-tuning is 
equivalent to choosing the asymptotic value of the dilaton to be small.

\section{Classification of continuous transitions}
\lab{class}
\subsection{General results}

In this section we investigate the role of the sub-leading
asymptotics in the scalar potential $V_{sub}(\f)$ in determining
the type of the phase transition. We shall not be completely
general, but only focus on two interesting cases where $V_{sub}$
is either an exponential or a power-law.

We state the results before detailing their derivation. We
consider two different classes of asymptotics:
\bea
  \mathrm{Case\,\, i:}\qquad V_{sub} &=& C~e^{-\kappa \f}, \quad \kappa>0, \qquad
  \f\to\infty\lab{case1}\\
  \mathrm{Case\,\, ii:}\qquad V_{sub} &=& C~\f^{-\a}, \quad \a>0, \qquad
  \f\to\infty\lab{case2}
\eea
Furthermore, we define the normalized temperature,
\begin{equation}\label{t}
  t=\frac{T-T_c}{T_c}.
\end{equation}
The asymptotics (\ref{case1}) and (\ref{case2}) the free energy density
difference,
\bea
   \mathrm{Order\,\,n\,\,transition:} \,\,\, \Delta F(t) &\sim& t^n,\qquad \mathrm{for}\,\, \kappa =
  \frac{\sqrt{\xi(d-1)}}{n-1},\,\, n\ge 2 \label{nth}\\
   \mathrm{Infinite\,\, order\,\, transition:} \,\,\, \Delta F(t) &\sim& e^{-A
  t^{-\frac{1}{\a}}},\qquad \mathrm{for}\,\,\a>0. \label{ath}
\eea
In particular when $\a=2$ one obtains a BKT type scaling in the
case (\ref{case2}):
\begin{equation}\label{BKT}
  \Delta F(t)\sim e^{-A t^{-\half}}, \qquad \a=2.
\end{equation}
Another special case is when the scalar field is the dilaton in
NCST where the normalization of the kinetic term in (\ref{action})
becomes $\xi = 4/(d-1)$. Then, in the case (\ref{case1}) above, one has an
nth order transition for $\kappa = 2/(n-1)$.

\subsection{Derivation in a coordinate invariant approach}
\lab{XYvar}

We would like to derive the results (\ref{nth}) and (\ref{ath}) in
an approach that is independent of the chosen coordinate system,
where one only needs to specify the asymptotics of the scalar
potential as in (\ref{case1}) and (\ref{case2}).  In this section
we introduce such an approach. The derivation of the thermodynamic
properties based on this approach is presented in the next
section.

Such a  coordinate independent approach was introduced in terms of
the so-called ``scalar variables'', in section 7 of
\cite{GKMN2}\footnote{See also the ``thermal superpotential''
method in \cite{GKMN2} that is equivalent to the scalar
variables.}. The idea is to introduce variables that transform as
scalars under the diffeomorphism that involve $r$ i.e. $r\to
\tilde{r}(r)$. Then, one reduces the complicated Einstein-scalar
system to a system of coupled first order equations for these
scalar variables.

One can always show that all physical results,
including the thermodynamic functions etc. will only depend on
this reduced system.

In case of a neutral black-hole in $d+1$ dimensions, one only
needs two such variables, that we call $X$ and $Y$:
\begin{equation}\label{XY}
  X(\f)\equiv \frac{\g}{d}\frac{\f'}{A'}, \qquad  Y(\f)\equiv
  \frac{1}{d}\frac{f'}{f~A'}.
\end{equation}
where the constant $\g$ is given by,
\begin{equation}\label{gam}
  \g = \sqrt{\frac{d~\xi}{d-1}}.
\end{equation}
These functions satisfy the following system of coupled first
order equations:
\bea\label{Xeq}
\frac{dX}{d\f} &=& -\g~(1-X^2+Y)\le(1+\frac{1}{2\g}\frac{1}{X}\frac{d\log V}{d\f}\ri),\\
\frac{dY}{d\f} &=& -\g~(1-X^2+Y)\frac{Y}{X }. \label{Yeq}
\eea
We leave the derivation of these and other equations below to App.
A. These equations are solved by imposing boundary conditions at
the horizon: Let us denote the value of the scalar at the horizon
as $\f_h$. Then regularity of the horizon fixes the values of
$X(\f_h)$ and $Y(\f_h)$ completely in terms of the scalar
potential:
\begin{eqnarray}
  Y(\f) &=& \frac{Y_h}{\f_h-\f} + Y_1+ \cO(\f_h-\f) \lab{Yhor},\\
  X(\f) &=& X_h + X_1(\f_h-f) + \cO(\f_h-\f)^2\lab{Xhor},
\end{eqnarray}
with
\begin{small}
\begin{equation}\label{detyh}
X_h = -\frac{1}{2\g}\frac{V'(\f_h)}{V(\f_h)}, \quad Y_h = -\frac{X_h}{\g}, \quad X_1 = -\frac{1}{4\g}\le(\frac{V''(\f_h)}{V(\f_h)} - \frac{V'(\f_h)^2}{V(\f_h)^2}\ri), \quad Y_1 = 1-X_h^2.
\end{equation}
\end{small}
All other sub-leading terms are determined similarly by perturbative expansion near $\f_h$.

Having determined the functions $X$ and $Y$ by solving these eqs.
 the  metric functions are obtained by the definitions (\ref{XY}) as:
\bea
A(\f) &=& A_0 + \frac{\g}{d}\int_{\f_0}^{\f} \frac{1}{X}d\tilde{\f},\lab{Aeq}\\
f(\f) &=&  \exp\le[\g \int_{\f_0}^{\f}
\frac{Y}{X}d\tilde{\f}\ri],\lab{feq} \eea
 where $\f_0$ is the boundary value of the scalar field, and in the second equation  we used the requirement that the metric function $f$ is unity
at the boundary.

This reduction corresponds to setting $\f$ as the radial variable,
with the hidden assumption that $r(\f)$ is a single valued
function, i.e.   $\f(r)$ is monotonically increasing.

Finally, one can invert (\ref{Xeq}) to express the scalar potential in terms of the scalar variables as,
\be\lab{V1} V(\f) = \frac{d(d-1)}{\ell^2} \le(1+Y-X^2\ri)
e^{-2\g\int_{-\infty}^{\f} (X(t)-\frac{Y(t)}{2X(t)})dt}. \ee

Derivation of thermodynamic relations in terms of the scalar variables is
presented in App. \ref{Apptherm}. Given the scalar potential $V(\f)$ one solves 
for $X(\f)$ in (\ref{Xeq}) and (\ref{Yeq}) obtains the thermodynamics of the 
corresponding BH solution from the following two equations.
\bea
\Delta F(\f_h) &=& \frac{1}{4G_D} \int^{\infty}_{\f_h} d\tilde{\f}_h~e^{(d-1)A(\f_h)}~\frac{dT}{d\tilde{\f}_h} ,\lab{Feq}\\
T(\f_h) &=& \frac{\ell}{4\pi(d-1)}~e^{A(\f_h)}~V(\f_h)~e^{\g
\int_{\f_0}^{\f_h} X(\f)~d\f}.\lab{Teq} \eea
The first one follows from the 1st law of thermodynamics (\ref{ent1}) and the expression for the entropy (\ref{ent3}). The second 
one follows from expressing (\ref{Trh}) in terms of the scalar variables. We refer to Appendix \ref{Apptherm} for details.

The constant $\ell$ that appears in these equations is the length scale, (\ref{relsinf}), that is given by
the curvature radius of the geometry at the boundary $\f_0$, which
is in turn determined by the value of the potential at $\f_0$:
\be\label{ell}
\ell^2 = d(d-1)\frac{1-X(\f_0)^2}{V(\f_0)}.
\ee
For example, in case of pure AdS, the scalar should be constant, hence  $X=0$ and from (\ref{ell}) follows the standard expression for the cosmological constant
 in d+1 dimensions.

 \section{Scaling of thermodynamic functions}
 \lab{scatherm}
\subsection{Free energy}

 Now, it is a matter of straightforward calculation to see how the free energy scales with temperature for asymptotics of the type (\ref{case1}) and (\ref{case2})
 that hold in the region $1\ll\f<\f_h$.
One first solves  $X$ and $Y$ from (\ref{Xeq}) and (\ref{Yeq}) and shows that,
\be\label{asXY} X(\f) = -\frac{1}{\sqrt{d}}+\delta X(\f), \qquad
Y(\f) = Y_0(\f) + \delta Y(\f), \ee
where $\delta X$ and $\delta Y$ are sub-leading with respect to the first terms in the 
limit $1\ll\f<\f_h$.

It turns out that we do not need the precise expressions for
$\delta X$, $Y_0$ and $\delta Y$  above to determine the type of the phase
transition. However they become important, for example, in
determining the viscosity coefficients around $T_c$. We present
them, in Appendix \ref{Apptherm}.

One determines the scale factor $A(\f)$
from (\ref{Aeq}) as,
\be\label{asA} A(\f_h) = -\frac{\g}{\sqrt{d}}\f_h
-\g\int^{\f_h}\delta X(\f)~d\f + \mathrm{const}. \ee
The constant piece is unimportant in what follows
and one can show that the second term is sub-leading with respect to the first one,
see Appendix \ref{AppC}

Substituting (\ref{asXY}) and (\ref{asA}) in (\ref{Teq}), one
finds that all the dependence on $\delta X$ cancels out and the
rate of approach of T toward $T_c$ is directly determined by
$V_{sub}$. Using the definition of the normalized temperature
(\ref{t}), one finds:
\begin{equation}\label{ast}
  t = V_{sub}(\f_h).
\end{equation}
Now the free-energy as a function of $t$ is given by (\ref{Feq}) and it can be rewritten as,
\begin{equation}\label{asF}
  \Delta F(t) \propto \int_0^t d\tilde{t}~e^{(d-1)A(\tilde{t})}.
\end{equation}
The dependence of the scale factor on $t$ should be found by
inverting (\ref{ast}) and substituting in (the leading term of)
(\ref{asA}): In the cases (\ref{case1}) and (\ref{case2}) one finds,
\bea
  \mathrm{Case\,\, i:}\qquad A(t) &=& \frac{1}{\kappa}\sqrt{\frac{\xi}{d-1}}\log(t/C) + \cdots, \qquad
  t\to 0^+ \lab{Acase1}\\
  \mathrm{Case\,\, ii:}\qquad A(t) &=& -\sqrt{\frac{\xi}{d-1}} \le(t/C\ri)^{-\frac{1}{\a}} + \cdots, \qquad
  t\to 0^+ \lab{Acase2},
\eea
The free energy follows from (\ref{asF}) as\\
\begin{tabular}{|p{12cm}|}
\hline \bea
  \mathrm{Case\,\, i:}\qquad \Delta F(t) &\propto & t^{\frac{\sqrt{\xi(d-1)}}{\kappa}+1}, \qquad
  t\to 0^+ \lab{Fcase1}\\
  \mathrm{Case\,\, ii:}\qquad \Delta F(t) &\propto & e^{C' t^{-\frac{1}{\a}}} t^{1+\frac{1}{\a}}, \qquad
  t\to 0^+ \lab{Fcase2},
\eea
\\
\hline
\end{tabular}
\newline
where $C'= \sqrt{\xi(d-1)}C^{\frac{1}{\a}}$ in the second
equation. We see that $\Delta F$ vanishes, as it should, for arbitrary but positive constants $\xi$, $\kappa$ and $\a$.

In the special case of
\be\lab{kappa} \kappa =\sqrt{\xi(d-1)}/(n-1) \ee
 in (\ref{Fcase1}), one obtains an nth order transition in case i. Moreover, if
 the scalar field is  dilaton, then $\xi$ is given by
(\ref{xincst}) and (\ref{kappa}) reduces to,
\be\lab{kappaNC} \kappa =\frac{2}{n-1}. \ee
In this case the required asymptotics in
(\ref{Vs}) for an order-n transition becomes,
\begin{equation}\label{Vsn}
  V(\f) \to e^{\frac{4}{d-1}\f}  + C
  e^{\frac{4n-2d-2}{(n-1)(d-1)}}, \qquad \f\to\infty.
\end{equation}
For example, for a 3rd order phase transition in a dual 2+1
dimensional theory, one should require,
\begin{equation}\label{Vsn33}
  V(\f) \to e^{2\f}  + C
  e^{\f}, \qquad \f\to\infty.
\end{equation}

On the other hand, the special case of $\a=2$ in (\ref{case2}) is
a BKT type transition where the free energy scales like\footnote{One can also get rid of the power-like corrections in (\ref{Fcase2}) by slightly modifying the
asymptotics in (\ref{case2}) by adding log corrections, but we shall not consider this here.}
$$F(t) \sim e^{\frac{-C'}{\sqrt{t}}}.$$
We note that the setting is rich enough to cover different BKT type
scalings that one encounters in the condensed-matter literature
for different values of $\a$. For example, in the dislocation
theory of melting in two-dimensional solids \cite{HalperinNelson}
one encounters a different value, $\a = 2.70541\dots$ in
(\ref{Fcase2}).

We also note that this set-up covers other, more general type of
{\em divergent  transitions} where one of the derivatives of the
free energy becomes not discontinuous, but diverges. For a recent
reference, cf. \cite{Janke}. Clearly, these type of transitions
are ubiquitous in the case i above, for arbitrary real $\kappa$.

\subsection{Fine-tuning the potential}
\lab{finetune}

Before continuing further, we would like to comment on the issue
of the transition region coinciding with the singular region in
Einstein's gravity, section \ref{singular}. In this paper we are only 
interested in how the thermodynamic quantities scale with the
reduced temperature $t$ and not in the overall size of these
quantities. In this case, one way to avoid the problem of
running into the singular region is to fine-tune the parameters of
the potential $V(\f)$ to parametrically separate the
singular region $\f_h\gg 1$ from the vicinity of temperatures
where the scaling sets in, i.e. $t\ll 1$.

Equation (\ref{ast}) suggests an obvious way to manage this
separation just by choosing the phenomenological parameter $C$ in
(\ref{case1}) and (\ref{case2}), small enough. In the Einstein frame how small it should
be chosen is determined by the demand that the size of the
curvature $R \ell^2$ is not too large\footnote{One can address the same issue in the string frame. 
In this frame the curvature invariants vanish in the transition limit (see Appendix \ref{AppRicci}), 
however a large value of the dilaton signals breakdown of our assumption of weak gravitational 
interactions. In section 6, we construct explicit examples and show that the condition derived here is essentially 
equivalent to choosing the boundary value of the dilaton small enough so that it stays small in the transition 
region.}. 
Suppose we decide to trust
Einstein's gravity up to the point where the curvature becomes some order 1
number  $A$. Then, in the region where $V$ is
approximately given by (\ref{Vs}), one has, \be\lab{ricci} R
\propto V_\infty~e^{2\sqrt{\frac{\xi}{d-1}}\f_h} \sim A/\ell^2,
\ee where the proportionality constants are $\cO(1)$. We refer the
reader to Appendix \ref{AppRicci} for a derivation of the
curvature invariants. Substituting this in (\ref{case1}) or
(\ref{case2}) determines how small the constant $C$ should be
chosen. In case 1, for example, one needs, \be\lab{Csm} C\ll \le(
\frac{A}{V_\infty\ell^2}\ri)^{\frac{\kappa\sqrt{d-1}}{2\sqrt{\xi}}}.
\ee In our case $C$ is a phenomenological parameter that can be
chosen arbitrarily small.

Another means to accomplish the separation of the transition region $t\approx 0$ and 
the singular region $\f_h \gg 1$ is to tune the boundary value of the dilaton to be very small. 
This can be done in an asymptotically AdS solution, see section 6.  

Either of these ways seem to be a crude way to avoid the problem and there is a prize that one pays. 
For example when we consider quantities given by ratios such as $\eta/S$ and $\zeta/S$, 
we will see that there is an ambiguity in determining their value at $T_c$.

\subsection{Entropy and specific heat}
\lab{scv}

As we focus on continuous transitions in this paper, the classical entropy difference vanishes
exactly at the transition\footnote{As discussed in section \ref{gens}, by vanishing of the entropy we really mean it becomes of $\cO(N^{-2})$
in the large N limit.}.
The scaling of the entropy near $T_c$ is determined by (\ref{ent3}) and (\ref{Acase1}), (\ref{Acase2}):
\bea
  \mathrm{Case\,\, i:}\qquad \Delta S(t) &\propto & t^{\frac{\sqrt{\xi(d-1)}}{\kappa}} + \cO(N^{-2}), \qquad
  t\to 0^+ \lab{Scase1}\\
  \mathrm{Case\,\, ii:}\qquad \Delta S(t) &\propto & e^{-C' t^{-\frac{1}{\a}}}+ \cO(N^{-2}), \qquad
  t\to 0^+ \lab{Scase2},
\eea
where $C'$ is defined below (\ref{Fcase2}). 

The specific heat is obtained from  $\Delta C_h = T d\Delta S/dT$ as,
\bea
  \mathrm{Case\,\, i:}\qquad \Delta C_v(t) &\propto & t^{\frac{\sqrt{\xi(d-1)}}{\kappa}-1}+ \cO(N^{-2}) , \qquad
  t\to 0^+ \lab{Cvcase1}\\
  \mathrm{Case\,\, ii:}\qquad \Delta C_v(t) &\propto & e^{-C' t^{-\frac{1}{\a}}} t^{-\frac{1}{\a}-1}+ \cO(N^{-2}) , \qquad
  t\to 0^+ \lab{Cvcase2},
\eea
Clearly, for any $\kappa> \sqrt{\xi(d-1)}$ the specific heat
diverges at the transition. In the special case of (\ref{kappaNC})
with $n=2$, it becomes discontinuous, as it should in a second
order transition.

\subsection{Speed of sound}\lab{speed}

The (normal) speed of sound is defined as  $c_s^2 = dP/dE$. In
isotropic systems that we  consider here, the pressure is given by
the free energy density and the speed of sound in the BH phase is most easily
determined from
\be\lab{ssoundBH}
c_{s,BH}^2 = \frac{S_{BH}}{C_{v,BH}}.
\ee

Now  we ask the question, what does $c^2_{s,BH}$ become in the limit $T\to T_c$? 
Equation (\ref{ssoundBH}) can be written as, \be\lab{ssbhn} c_{s,BH}^2 = \frac{ \Delta S +
S_{TG}}{\Delta C_v + C_{v,TG}}. \ee 
For the case of a 3rd or higher order transition, the leading order terms $\Delta S$ and $\Delta C_v$
vanish as in section \ref{scv} and one is left with the terms of $\cO(N^{-2})$ both in the
numerator and the denominator. Their ratio is $\cO(1)$ and it is given by $S_{TG}/C_{v,TG}$ which is 
nothing but the speed of sound in the thermal gas phase. Therefore the speed of sound stays 
continuous across the transition, which is indeed what should happen for a transition of order higher than 2. 
In the case of a second order transition $\Delta S$ vanishes, but $\Delta C_v$ stays constant and one finds a 
discontinuity in the speed of sound. All of this is completely in accord with a system that 
obeys the laws of thermodynamics .  However there is an important ambiguity which we discuss below. 

\subsection{An order of limits issue and divergence in the dilaton}
\lab{ool}

An ambiguity arises when we consider  the $N\to\infty$ limit in the 
discussion above: We observed that taking the $t\to 0$ limit while keeping $N$ finite one obtains $c_s^2 \sim \cO(1)$ 
at the transition point. Now, one can further take the $N\to\infty$ limit and $c_s^2$ still stays the same. On the other 
hand, had we taken the $N\to\infty$ limit first, then the  $\cO(N^{-2})$ terms in (\ref{ssbhn}) would drop out and we would obtain 
the result $c_s^2 \sim t$ around the transition. Thus, we learn that the limits $N\to\infty$ and $t\to 0$ do not commute. 

The resolution of the puzzle is clear: one cannot really take the $t\to 0$ limit for finite $N$ without taking into account the 
gravitational corrections. This is because the Einstein's theory breaks down in this limit. For definiteness, 
let us consider Einstein's gravity as an effective theory to a $d+1$ dimensional non-critical string theory\footnote{The precise nature
of this theory is irrelevant  for the sake of  discussion. One could also consider a compactification of II(A)B on
a $9-d$ dimensional  internal manifold.}. Then, the effective Newton's constant  at point r is given by
\be\lab{ges} G_{eff} = (16\pi M_p^{d-1}N^2)^{-1} e^{2\f(r)} \sim
g_s^2 \ell_s^{d-1} e^{2\f(r)}, \ee where $g_s$ is the value of the dilaton field on the boundary. To be definite, we consider an asymptotically AdS
geometry where the dilaton approaches a constant value $\f(r)\to \f_0$
as $r\to 0$ and $g_s= \exp(\f_0)$. 
This  implies that, for finite N there is
a limit in $r_h$ beyond which we cannot ignore the gravitational interactions.
In the specific case of (\ref{case1}) it is determined from (\ref{ges}) as\footnote{ we considered the
dimensionless effective Newton's constant $G_{eff}\ell^{d-1}$.}
\be\lab{tlim} t_{lim} \sim C (4\sqrt{\pi} (M_p\ell)^{(d-1)/2}N)^{-\kappa}, \ee where $C$ is the
constant defined in (\ref{case1}). Beyond this limit one should take
into account the gravitational corrections.

One can ask whether one can send $t_{lim}$ to zero by fine-tuning a parameter in the theory. One  can 
indeed fine-tune $t_{\lim}$ to be arbitrarily small by choosing $C$ or $1/N^2 \sim e^{\f_0}$ to be small. 
Sending $C$ to zero is not desirable, as in this case the subleading term in  (\ref{Vs}) vanishes and 
the resulting potential displays no transition. But  one can send $e^{\f_0}$ to zero instead. This corresponds to 
the strict $N\to\infty$ limit. 

The conclusion is that for a system that corresponds to finite $N$, regardless how large N is, one should consider $\cO(G_D)$ corrections in order to calculate the value of a quantity 
such as $c_s^2$. On the other hand, the expected result that $c_s^2$ becomes discontinuous at $T_c$ in a second order 
and continuous  in a higher order transition should still follow after taking into account the $\cO(G_D)$ corrections 
because $\Delta S$ and $\Delta C_c$ should still vanish at $T_c$ also after the corrections. The only ambiguity is 
in determining the value of a quantity given by a {\em ratio} such as $c_s^2$, precisely at $T_c$. We encounter 
other examples of such quantities below.  

%
%

\section{Viscosities and dispersion}
\lab{viscdisp}

Viscosity coefficients are important observables probing the
dissipative behavior  in the dual field theory at finite
temperature. In our case there are only two viscosity
coefficients, that determine the response of the system to {\em
shear} and {\em bulk} deformations respectively.

\subsection{Shear deformations}

The shear viscosity is given on the gravity side by the flux  of
the ``shear'' gravitons, i.e. the traceless-transverse component
$h_{TT}$  of the metric fluctuations. Due to the well-established
universality in the case of two-derivtive gravity
\cite{PSS}\cite{BuchelLiu} the shear viscosity to entropy ratio in
the BH phase is given by,
\be\lab{shearent}
\frac{\eta}{S} = \frac{1}{4\pi} + \cdots
\ee
where the ellipsis refer to $\a'$ and $1/N$ corrections. We note
that at the classical level in the two-derivative Einstein-scalar
gravity no temperature dependence arises in this ratio, and the
result is simply given by (\ref{shearent}) for at any $T\ne T_c$.

What happens exactly at $T=T_c$ is similar to the story of the
speed of sound above. To be more precise by including  the $1/N^2$
corrections, one should write $\eta = \eta_0 + \delta \eta$ where
$\eta_0$ is the leading order piece in the $1/N^2$ expansion,
$\eta_0 = \Delta S/4\pi$ and $\delta \eta$ refers to $1/N^2$
corrections. One has,
\be\lab{shearent2}
\frac{\eta}{S_{BH}} = \frac{\eta_0 + \delta\eta}{\Delta S + S_{TG}} = \frac{1}{4\pi} + \cdots
\ee
 In the limit $T\to T_c$ the leading order terms vanish, but one still obtains a $\cO(N^0)$ answer,
\be\lab{shearent3} \frac{\eta}{S_{BH}} \to
\frac{\delta\eta}{S_{TG}}. \ee
In order to determine the value of this, one has to include the $\cO(1/N^2)$ corrections in the analysis. 
The situation is different in the strict $N\to\infty$ limit where the $\cO(N^{-2})$ pieces in (\ref{shearent2}) 
drop out and one retrieves  the universal value $\eta/s = 1/4\pi$ also at the transition point. 

\subsection{Bulk deformations}
\lab{bulkdeform}

The {\em bulk viscosity} on the other hand is more interesting as
it displays T dependence already at the two-derivative level
\cite{GubserBulk}\cite{GKMinN}. It is given on
the gravity side, by the flux of the rotationally-invariant
fluctuations of the metric $h_b\equiv h_{11} = h_{22} = \cdots =
h_{d-1,d-1}$. The computation is much more involved than the shear
case, due to mixing of these fluctuations with the fluctuations of
the scalar field $\delta\phi$. An ingenious method to disentangle
this mixing is devised by Gubser et al in \cite{GubserBulk}. The
main idea is to consider a gauge where the fluctuations of the
scalar is  set to zero. This is equivalent to carrying out the
computation by using $\phi$ as the radial variable. The method is
described in detail in \cite{GubserBulk} for the case of $d=4$. Here, we generalize this
calculation to arbitrary dimensions. This is straightforward, yet technically involved and we present the details
of this computation in App.   \ref{AppF} .
We shall only highlight the main steps and results here.

Kubo's linear response theory \cite{Kubo} relate the bulk viscosity to the retarded Green function of the stress tensor as
\footnote{The subscript $0$ refers to the leading term of the quantity in the $1/N^2$ expansion.} 
\be\lab{Kubozeta}
\zeta_0 = -\frac{4}{(d-1)^2}\lim_{\o\to0} \frac{1}{\o} Im\, G_R(\o)
\ee
where the Green function is evaluated at zero momentum and it is defined by,
\be\lab{Green}
G_R(\o) =   -i \int dt\, d^{d-1}x e^{i\o t} \theta(t) \la [\half T^i_i(t,\vec{x}), \half T^j_j(t,\vec{x})]\ra,
\ee
and $i,j$ denote the spatial components only. The imaginary part
of the Green function maps to the flux of the $h_b$ gravitons on
the gravity side. It is determined by a long and non-trivial
computation (cf. \cite{GubserBulk} and App. \ref{AppF} of this
paper). The computation is carried out in the gauge $\delta\f =0$.
In other words $\f$ is taken as the radial variable. We denote the
metric functions in this gauge as,
\be\lab{metphi}
ds^2 = e^{2A(\f)} \le(-f(\f) dt^2 + dx_{d-1}^2\ri) + e^{2B(\f)} \frac{d\f^2}{f(\f)}.
\ee
The final result reads,
\be\lab{hbflux}
Im\, G_R(\o) =  -\frac{{\cal F}}{16\pi G_D} = -i \frac{\xi}{64 \pi G_D} \frac{f}{A^{'2}} e^{(d-2)A-B} \le(h_b^*h_b' - h_b h_b^{*'}\ri)\bigg|_{\f_h}
\ee
where prime denotes $d/d\f$. The metric fluctuation $h_b$ satisfies the following second order equation :
\be\lab{fluceq1}
h_b'' +h_b' \le( \frac{2\xi}{d-1}\frac{1}{A'} + (d-2)A' - 3B' + \frac{f'}{f}\ri) + h_b\le( \frac{\o^2}{f^2} e^{2B-2A} + \frac{f'}{f}B' - \frac{\xi}{d-1}\frac{f'}{f A'}\ri)=0.
\ee
Here prime denotes $d/d\f$. In passing we note that this equation  looks simpler and more transparent in the $r$-variable:
\be\lab{fluceq2}
\ddot{h}_b +\dot{h}_b \le((d-1)\dot{A}  + \frac{\dot{f}}{f} + 2 \frac{\dot{X}}{X}\ri) + h_b\le( \frac{\o^2}{f^2}-  \frac{\dot{f}}{f}\frac{\dot{X}}{X} \ri)=0.
\ee
where dot denotes $d/dr$ now. We note that reference to the normalization of the kinetic term $\xi$ disappears in this form.
We also note that it reduces to the shear fluctuation equation  for $X=const.$

Equation  (\ref{fluceq2}) is solved by the boundary conditions $h_b = 1$ on the boundary
and the in-falling boundary condition at the horizon:
\be\lab{infall}
h_b \to c_b (r_h- r)^{-i\frac{\o}{4\pi T}},\qquad r\to r_h.
\ee
Here $c_b$ is a T-dependent multiplicative factor. Inserting (\ref{infall}) in (\ref{hbflux}) (and changing variables to $\f$) yield,
\be\lab{hbflux2}
Im\, G_R(\o) =  -\frac{{\cal F}}{16\pi G_D} = \frac{\o (d-1)^2}{32 \pi \xi} S_{BH} |c_b|^2 \le( \frac{V^{'2}}{V^2}\ri)\bigg|_{\f_h}
\ee
Now, the bulk viscosity follows from the Kubo's formula (\ref{Kubozeta})
\be\lab{visc1}
\frac{\zeta_0}{S_{BH}} = \frac{1}{8\pi \xi}|c_b|^2 \le( \frac{V^{'2}}{V^2}\ri)\bigg|_{\f_h}.
\ee

\subsection{Universality at $T_c$}

In the limit $\f_h\to\infty$ i.e. $T\to T_c$ one obtains {\em a universal result for the bulk-viscosity-to-entropy ratio}. 
As shown in \cite{GKMinN} the coefficient $|c_b|$ approaches to 1 in that limit. The ratio $V'/V$ also becomes a constant that
is given by the exponent in  equation  (\ref{Vs}). Thus, (\ref{visc1}) becomes,
\be\lab{visc2}
\frac{\zeta_0}{S_{BH}}\bigg|_{T_c} = \frac{1}{2\pi (d-1)},
\ee
at the transition. We note that this is a {\em universal result}
in the sense that it  depends neither on the value of $T_c$ nor on
the details of the underlying field theory system, nor on the
order of the phase transition. 
The details of the  computation that leads to this result are presented in App \ref{AppF}.

As for the shear case, the result is valid {\em only in the limit where one can ignore gravitational interactions.} 
Just as in the discussion below equation (\ref{shearent}) above, if there exists $1/N^2$ corrections 
then the result changes not by $\cO(N^{-2})$ but by $\cO(N^0)$. 
Denoting the
$1/N^2$ correction to $\zeta_0$ as $\delta \zeta$, one has,
\be\lab{visc22}
\frac{\zeta}{S_{BH}} = \frac{\zeta_0 + \delta\zeta}{\Delta S + S_{TG}} \to \frac{\delta\zeta}{S_{TG}}, \qquad as\,\,\, T\to T_c.
\ee
The latter is determined by computing the $1/N^2$ corrections in
the background and there is no reason that it should coincide with
(\ref{visc2}), even in the large $N$ limit. However if one considers a system with $N=\infty$ strictly, then 
one should take the $N\to\infty$ before $t\to 0$ and the result can be trusted in principle. In the next section 
we construct explicit kink solutions that flow from an AdS on the boundary (with a fixed value of $e^{\f_0}$) 
and linear-dilaton in the deep interior. In these solutions one can indeed send $e^{\f_0}\to 0$ in a consistent fashion
and one obtains the universal result above. It would be extremely interesting to find a similar solution in IIB and check 
whether the result given by (\ref{visc2}) holds there.


\subsection{Scaling of viscosities near $T_c$}

There is no ambiguity however, in the scaling of the viscosities $\eta$ and $\zeta$ near $T_c$, as this is determined by the leading piece in
the $1/N^2$ expansion. From the results above, we find that
both $\eta$ and $\zeta$ scales as the entropy density near $T_c$:
\begin{equation}\label{ez}
  \eta,\, \zeta \propto \Delta S(t) \propto e^{(d-1)A(t)}.
\end{equation}
The latter is given by (\ref{Acase1}) and (\ref{Acase2}) in cases
i and ii:
\bea
  \mathrm{Case\,\, i:}\qquad \eta,\, \zeta &\propto& t^{\frac{\sqrt{\xi(d-1)}}{\kappa}}\qquad
  t\to 0 \lab{ezcase1}\\
  \mathrm{Case\,\, ii:}\qquad \eta,\, \zeta &\propto& e^{C' t^{-\frac{1}{\a}}} , \qquad
  t\to 0 \lab{ezcase2},
\eea
where $C'$ is a constant given below equation  (\ref{Fcase2}). The
latter is the typical scaling of the order parameters in the
Kosterlitz-Thouless transition \cite{BKT}, \cite{Kosterlitz}.

\subsection{Dispersion relation of pressure waves}

Now, we are in a position to write down the dispersion relation of pressure waves near criticality, including dissipative terms. The hydrodynamics
predict a dispersion relation of the form,
\be\lab{dispersion}
\omega = c_s {\bf q} - \frac{i}{T} \le(\frac{d}{d-1}\frac{\eta}{S_{BH}} + \half \frac{\zeta}{S_{BH}}\ri) {\bf q}^2 +\cdots
\ee
From the results in the previous two subsections, eqs. (\ref{shearent}) and (\ref{visc2}), {\em in the strict $N=\infty$ limit} one finds,
\bea
  \mathrm{Case\,\, i:}\qquad\omega &\to&  c_s ~{\bf q} - \frac{i}{4\pi T_c}\le(\frac{d+1}{d-1}\ri){\bf q}^2, \qquad
  t\to 0 \lab{cscase1}\\
  \mathrm{Case\,\, ii:}\qquad\omega&\to& c_s~{\bf q} - \frac{i}{4\pi T_c}\le(\frac{d+1}{d-1}\ri){\bf q}^2, \qquad
  t\to 0 \lab{cscase2},
\eea
where we only show the leading terms in $t$ in both of the terms. One should remember the ambiguities in these formula 
for finite N as explained in the previous subsection.

\section{UV geometry and  analytic kink solutions}
\lab{ansol}

So far we have not specified the geometry near the boundary. The only assumption we made in the derivation of section 2 is that $A'(r)<0$ near $r\approx 0$. This includes the case of asymptotically AdS backgrounds. In fact, for holographic applications this is the most desirable case. Therefore it would be interesting to find explicit 
solutions that flow from an AdS geometry in the UV to a geometry of the type given by the solution of the IR potential (\ref{Vs}) with 
$V_{sub}$ given by either of (\ref{case1}) or (\ref{case2}). In this section we construct such explicit examples for the case of 
(\ref{case1}). This will the most interesting case for applications to holographic superfluids in \cite{exotic2}.  

An asymptotic AdS geometry is also desirable for another reason. In the case when the scalar can be regarded as the dilaton of 
NCST, its value goes over to a constant   $\f(0)=\f_0$  in an asymptotic AdS  geometry which provides a tunable parameter. On the other hand, the effective Newton's constant is related to this value as,
\be\lab{effnew}
G_D \propto e^{2\f_0}.
\ee
Therefore, this value can be tuned to make the gravitational interactions very small. In other words, $\f_0$ will be the parameter that corresponds 
to the ``color" N in a dual field theory 
\be\lab{effnew1}
N^2 \propto e^{-2\f_0}.
\ee
The large N limit corresponds to tuning $\f_0$ to be very large. 

At the same time, we argued in section \ref{finetune} that in order to keep the transition region  away from the curvature singularity, 
one needs another fine tuning $C\ll 1$ where $C$ is the constant defined in (\ref{case1}). In principle, these two separate fine-tunings 
may contradict each other. Here, we will show by  constructing explicit examples that, on the contrary, these two fine-tunings are essentially 
the same\footnote{This is not hard to see directly from (\ref{case1}) where shifting the vev of $\f$ can be regarded as tuning the constant $C$.}.  We shall focus on the case where the scalar field is regarded as the dilaton and its normalization in (\ref{action}) is given by 
$\xi = 4/(d-1)$. 

\subsection{Two-exponent potentials}

For notational simplicity we define the function $\l$
\be\lab{lambda}
\l \equiv e^{\f}, \qquad  \l_0 \equiv e^{\f_0}.
\ee
Consider the potential, 
\be\lab{dilpot1}
V(\l) = V_\infty \l^{\frac{4}{d-1}}\le(1+ C \l^{-\ka} \ri).
\ee 
According to the discussion in section \ref{class}, the black-hole solution that follows from this potential exhibits a continuous type 
Hawking-Page transition with the order of the transition determined by the constant $\kappa$ as  in (\ref{kappaNC}). For example 
the transition is second order for $\kappa=2$.  The transition region is in the asymptotic IR region $\l_h \gg 1$. 

We observe that this simple potential allows for an AdS extremum at\footnote{We assume $C>0$.} 
\be\lab{min1}
\l_0 = C^{\frac{1}{\ka}} \le(\frac{\kappa(d-1)}{4}-1 \ri)^{\frac{1}{\ka}}
\ee
when $\ka (d-1) >4$. As a result, {\em the two-exponent potential (\ref{dilpot1}) should have a kink solution that flows from an asymptotically AdS geometry in the UV at $\l=\l_0$ towards the 
asymptotically linear-dilaton geometry in the IR at $\l=\infty$.} 

We observe from (\ref{min1}) that the large N limit (\ref{effnew1}) corresponds to $C\ll 1$. This is reassuring because the same limit 
is required in section \ref{finetune} in order to separate the large curvature region from the transition region. 

One can construct this solution analytically for the thermal gas case (\ref{TG}),  for special values of $\ka$ and $d$. 
This can easily be done by the method of scalar variables,  Appendix \ref{AppA}\footnote{It can also be obtained by the method of ``fake super-potential'', \cite{Dan}.} as follows: The TG solution corresponds to the special solution of (\ref{Yeq}) $Y=0$. In this case the Einstein-dilaton system reduces to a single first order differential equation, 
\be\lab{sinf}
\frac{dX_0}{d\f} = -\g~(1-X_0^2)\le(1+\frac{1}{2\g}\frac{1}{X_0}\frac{d\log V}{d\f}\ri),
\ee
where $\g$ is given by (\ref{gam}),
\be\lab{gam2}
\gamma = \frac{2\sqrt{d}}{d-1}.
\ee
A very nice coincidence is that one can construct an analytic solution to (\ref{sinf}) precisely in the most interesting case, namely 
$\kappa=2$, $d-1=3$. This case corresponds to a second order phase transitions in three spatial dimensions, indeed the most relevant case 
for many condensed matter applications!  \cite{exotic2}. The solution reads, 
\be\lab{Xsol} 
X_0(\l) = -\half \frac{\l^2-\l_0^2}{\l^2+\half\l_0^2}.
\ee 
In the UV limit $X_0\to 0$ which indeed corresponds to an asymptotically AdS geometry\footnote{Equation (\ref{sinf}) looks singular in the limit $X_0\to 0$. In fact it is not as the pole of $X_0$ at $\l_0$ is cancelled against the pole in $V'(\l)$}. To see this, one  recalls the definition of $X_0$ in (\ref{XY}) and vanishing $X$ corresponds to constant dilaton. 

One can construct the metric and the dilaton in the r-frame explicitly by solving equations (\ref{Ap}) and (\ref{fp}) with $X=X_0$ given by (\ref{Xsol}) and $Y=0$. The result is, 
\bea
\lab{metsol}
ds^2 &=& \l_0^{-\frac43}~\frac{\ch^{\frac23}(\frac{3r}{2\ell})}{\sh^{2}(\frac{3r}{2\ell})}\le(dt^2 + dx_{d-1}^2 +  dr^2\ri),\\
\lab{lamsol}
\l(r) &=&  \l_0~\ch(\frac{3r}{2\ell}).
\eea
Clearly, the geometry becomes AdS near $r=0$. The metric in the string-frame is obtained
from (\ref{metsol}) by the Weyl transformation $ds^2_{st} = \l^{4/3}ds^2 $: 
\be
\lab{metsolst}
ds_{st}^2 = \coth^2(\frac{3r}{2\ell})\le(dt^2 + dx_{d-1}^2 +  dr^2\ri).
\ee
In the IR region $r\to\infty$ the background in the string-frame becomes linear-dilaton:
\be
\lab{metlin}
ds_{st}^2 \approx dt^2 + dx_{d-1}^2 +  dr^2, \qquad \f(r) \approx \frac{3r}{2\ell}.
\ee
Therefore the background (\ref{metsolst}) and (\ref{lamsol}) is a solution to the 5D Einstein-dilaton 
theory that interpolates between an AdS background near the boundary and a linear-dilaton background in the deep interior 
in the string frame. 

It would be very useful to obtain an analytic black-hole solution to the potential (\ref{dilpot1}) by turning on a blackness function in (\ref{metsol}). In the language of scalar variables (\ref{Xeq}), (\ref{Yeq}), this corresponds to turning on a non-trivial $Y$. Now, the system becomes more complicated 
and unfortunately we could not find an analytic solution. However a numerical solution can easily be constructed by solving the system
by a numerical code. We checked that the black-hole obtained this way obeys the general conditions described in  section \ref{condts}, 
in particular it asymptotes the TG kink above in the limit $r_h\to\infty$. To our knowledge, the closest analytic BH solutions are the Chamblin-Reall  
solutions to single exponential potentials \cite{Chamblin}. However, these solutions do not exhibit any Hawking-Page transitions. 

\subsection{Three-exponent potentials}

The analytic solution found above is for the particular case $\ka=2$, $d-1=3$. It is interesting to search for similar analytic solutions for
an arbitrary $\k$ and $d$. This is not possible  with a simple two-exponential dilaton potential. Instead we consider a three-exponential 
potential of the form:
\be\lab{dilpot2}
V(\l) = V_\infty \l^{\frac{4}{d-1}}\le(1+ C \l^{-\ka} +\tilde{C} \l^{-2\ka}\ri).
\ee 
This potential possesses AdS extrema for generic values of $\ka$ and $d$.  
Let us again consider a thermal-gas type \ref{TG} ansatz. One can find an analytic solution for the special value of $\tilde{C}$:
\be\lab{C2}
\tilde{C} = \frac{C^2}{(\ka+2)^2}\le(1+\ka -\frac{\ka^2}{4}(d-1)\ri).
\ee
In fact this class of potentials can be regarded as a generalization of (\ref{dilpot1}) with $\ka=2$ and $d-1=3$. In this special 
case the constant $\tilde{C}$ in (\ref{C2}) vanishes and the three-exponential potential (\ref{dilpot2}) reduces to (\ref{dilpot1}). Solution 
of (\ref{sinf}) in the general case reads, 
\be\lab{Xsol2} 
X_0(\l) = -\frac{1}{\sqrt{d}} \frac{\l^\ka-\l_0^\ka}{\l^\ka+\frac{1}{q}\l_0^2}, \qquad q\equiv  \frac{\ka}{2}(d-1)-1.
\ee 
One checks that, this solution reduces to (\ref{Xsol}) in the special case of $\ka=2$, $d-1=3$.  The special point 
$\l_0$ corresponds to the extremum of (\ref{dilpot2}) and it is given by, 
\be\lab{min2}
\l_0 = \le(\frac{C q}{\ka+2}\ri)^{\frac{1}{\ka}}.
\ee
The solution (\ref{Xsol2}) corresponds to a kink that interpolates between an AdS geometry at $\l_0$ to an asymptotically 
 linear-dilaton background  at $\l\to\infty$. 
 
 We again observe that by tuning the constant $C$ to small values one can establish two goals at the same time, namely to separate the transition region $t\approx 0$ and the high curvature region $\l_h\gg 1$ and to make the effective 
 gravitational coupling weak.

It is not possible to obtain a closed form expression for the metric in the r-frame
 \footnote{At best one can express r in terms of $\l$ as a combination of hypergeometric functions.} for generic $\ka$ and $d$ 
 but it is easy to write down the metric directly in the $\l$-frame: 
\be\lab{lsol}
ds^2 = \l^{-\frac{4}{(d-1)}}\le(1-\frac{\l_0^\ka}{\l^\ka}\ri)^{-\frac{2}{q}}\le[dt^2+dx_{d-1}^2
+ c_q \le(1-\frac{\l_0^\ka}{\l^{\ka}}\ri)^{\frac{2}{q}-2}\l^{-2}~d\l^2  \ri],
\ee
 where $q$ is given in terms of $\ka$ and $d$ in (\ref{Xsol2}) and the constant $c_q$ is given by,
 \be\lab{cq}
c_q=  \le(\frac{2\ell}{q(d-1)}\ri)^2.
 \ee
  The metric is defined in the range $\l\in(\l_0,\infty)$. 
 It is easy to see that the metric becomes AdS near $\l\approx \l_0$ by a change of variable $\l-\l_0 \propto r^q$ near $r\approx 0$. 
 On the other hand, for large values of $\l$ 
 it asymptotes to the linear-dilaton background in the string frame:
\be
\lab{metlin2}
ds_{st}^2 \approx dt^2 + dx_{d-1}^2 +  \le(\frac{2\ell}{q(d-1)}\ri)^2~d\f^2.
\ee 
The solution given by (\ref{Xsol2}) constitutes a one-parameter family 
of kink solutions---parametrized by $\ka$---that interpolates between AdS and linear-dilaton backgrounds 
in the string-frame. The geometry is regular everywhere in the string frame.  
It would be very interesting to embed this geometry in IIB. One may be able to do so, by finding an analogous 
solution in a consistent truncation of IIB with non-trivial dilaton.  
Apart from our main interest in this work, namely that  black-holes on these backgrounds exhibit continuous  type Hawking-Page transitions 
with the order parametrized by $\ka$, they may have other applications in different  contexts.

\section{Stringy corrections}
\lab{stcor}

In this paper, we consider two-derivative Einstein-scalar gravity in the limit of small gravitational
corrections, $G_D\to 0$. On the other hand, we showed that the continuous phase change occurs in a
singular limit, where
the black-hole horizon {\em marginally traps} a curvature
singularity.  A crucial issue then, is the possible  $\a'$
or $G_D$ corrections at this point. We comment on these issues in this section.

First of all, we should stress that the results that are obtained
above are consistent and robust in the framework of Einstein's
gravity. The two-derivative theory breaks down only near curvature
singularities, or when the dilaton field becomes very large. 
On the other hand the curvature singularity is always
cloaked by the BH horizon as one approaches the transition point
$t\to 0^+$, and the dilaton  can always be kept from growing too large as long as $t\neq 0$
 strictly by fine-tuning the dilaton potential. How to manage this is explained in
section \ref{finetune}. Therefore the questions, as well as the
answers that are studied so far are definite within this
framework.

On the other hand, a basic motivation for us to study continuous
phase transitions in gravity concerns possible applications in the
dual gauge theories or condensed matter systems through the
holographic correspondence, and a robust framework to relate the
two sides of the correspondence is string theory rather than
Einstein's gravity. Therefore, one should consider possible
stringy corrections if one desires to interpret the results of
this paper from an holographic point of view.

\subsection{Higher derivative corrections}

The stringy higher derivative corrections to Einstein's gravity are measured by the typical curvature of the geometry
$R\ell_s^2$  in units of string length $\ell_s = \sqrt{\a'}$. Concerning these corrections, one can ask two separate questions:

\begin{enumerate}
\item Are the conditions derived in sections \ref{condts}
and \ref{class} for the existence and the type of the continuous
transitions still valid once we include the $\a'$ corrections?

\item For which quantities and in which regions can we trust the observables  if we ignore the $\a'$ corrections?
\end{enumerate}

Let us consider the first question above. For existence of a
continuous phase transition, clearly, the ($\a'$ corrected)
geometries should asymptotically become the same in the IR region.
For a BH geometry that asymptotes a TG geometry near the boundary,
{\em this condition can most generally expressed as vanishing of
the ADM mass of the BH.}

Let us now specify to the case when the scalar is the dilaton
field of the $d+1$ dimensional non-critical string theory. Then
the normalization of the kinetic term is given by (\ref{xincst}).
The required asymptotics for a continuous phase transition is
given by (\ref{Vsncst}):
$$ V \to e^{\frac{4}{d-1} \f}.$$
In the string frame $g_{s,\m\n} = \exp(2\f/d-1) g_{\m\n}$ the
action reads
 \be\lab{staction}
 {\cal A}_s = \frac{1}{16\pi G_D} \int d^{d+1}x \sqrt{-g_s}e^{-2\f} \le( R_s - \xi_s (\6\f)^2 +
 V_s(\f)+\cdots
  \ri),
\end{equation}
where the subscript s refer to string frame quantities and the
ellipsis denote the (unknown) higher derivative corrections. Here
the string frame potential is related to the Einstein frame
potential by, \be\lab{Vse} V(\f) = V_s(\f) e^{\frac{4}{d-1}\f}.
\end{equation}
Therefore  the requirement for existence of a continuous
transition now becomes $V_s\to const.$ in (\ref{staction}), as
$\f\to\infty$. The remarkable fact is that, the unique solution to
this asymptotic potential with the TG ansatz (or with the BH
ansatz in the limit $\f_h\gg 1$) is nothing else but the
linear-dilaton solution, that is $\a'$ exact! Therefore, we
expect that the condition (\ref{Vs}) be $\a'$ exact in
this case.

There is still a possibility of $\a'$ corrections in the
sub-leading terms (\ref{case1}) and (\ref{case2}). This means that
the constants $\kappa$ and $\a$ in these equations can get
renormalized by the $\a'$ corrections\footnote{There is also the
possibility of power-law corrections in (\ref{case1}) and
log-corrections in (\ref{case2})}. However, clearly this would
only affect the definition of these constants. In other words, an
nth order transition would be given by $\kappa = 2/(n-1)
+\cO(\a')$ instead of (\ref{kappaNC}).

Finally, in the case when the scalar is {\em not} the dilaton of a
non-critical string, then there is a possibility that also the
leading term (\ref{Vs}) be corrected. Avoiding this is another
benefit of identifying the scalar with the dilaton of $d+1$
dimensional non-critical string theory.

Now we move on to the second question above. First consider the
thermodynamic functions: The entropy of the black-hole should
receive the usual $\a'$ corrections. These are of two different
origin: a) The definition of the entropy as the horizon size will
be corrected by Wald's generalization \cite{Wald}; b) the
background geometry itself will receive $\a'$ corrections. The
definition of the surface gravity, hence the temperature, is
subject to similar corrections of two separate origin. However,
the first law of thermodynamics should still be valid, therefore
we expect that the equation (\ref{Feq}) (or (\ref{F})) should
survive the corrections. Clearly, all of the other
thermodynamic quantities that we obtained in section
\ref{scatherm} are subject to $\a'$ corrections in the region of
high curvature in the Einstein frame. We expect these corrections to renormalize constants $\kappa$ and
$\a$ in (\ref{case1}) and (\ref{case2}). 

Secondly, one can ask if the thermodynamic quantities are subject
to higher derivative corrections away from the transition region.
The size of the corrections away from the singularity should be
determined by the ratio $\ell_s/\ell$ where $\ell$ is the typical
size of the asymptotic geometry, e.g. it is the AdS radius, if the
near boundary geometry is asymptotically AdS. This ratio is
holographically  related to an observable in the  dual field
theory. For instance, it is proportional to the gauge string
tension in the holographic  QCD models \cite{GKN}, whereas it is
related to the interaction strength between the spins in
spin-models \cite{exotic2}. Then, the issue of corrections to
thermodynamic quantities become a phenomenological question, and
the answer depends on the particular dual field theory that one
desire to describe holographically. For example, in the
holographic QCD models considered in \cite{GKN}, it was found by
matching the string tension on the lattice-QCD that the ratio is
$\cO(10^{-1})$ and the $\a'$ corrections are small.

Finally, we note that the  quantities that are naturally evaluated
in the string frame should be trustable in the vicinity of the
transition because the curvature invariants in the string frame in the transition region becomes very small 
as shown in Appendix \ref{AppRicci}. Such quantities involve F-string and $D-brane$
configurations in the probe approximation. In the companion paper,
\cite{exotic2}, we map these configurations to spin-correlation
functions in a dual spin-model. Therefore, we expect that these
important observables should be immune to $\a'$
corrections---except that corrections may arise through dependence
on the scaling constants $\kappa$ and $\a$ in (\ref{case1}) and
(\ref{case2}). We refer the reader to \cite{exotic2} for a more
detailed discussion.

\subsection{$1/N^2$ corrections}

Similarly the gravitational
interactions for finite $G_D$ also becomes important near the
transition region $t\sim 0$. This is because the dilaton
field\footnote{One should consider the case when the scalar is not
dilaton separately, see the discussion in section \ref{discuss}.}
diverges precisely at $t=0$. This means that the effective string
coupling $g_s$ becomes large in this region, unless $N$ can be set
to $\infty$ in a strict manner.

If in the dual field theory one has finite $N$---regardless how
large it is---our calculations can only be trusted in the region
\be\lab{tlim2} t > C (4\sqrt{\pi} (M_p\ell)^{(d-1)/2}N)^{-\kappa}.
\ee For a derivation of this bound, we refer to section \ref{ool}.
Beyond this point, one should take into account corrections.
Therefore, in the case of large but finite $N$ (small but finite
$G_D$), what one has to do in order to compute the rational
quantities at $T_c$, is to compute  the spectrum of fluctuations
in the TG geometry, and from this compute the determinant
correction in the saddle-point approximation, in the limit $T\to
T_c$.

We avoided this problem here by tuning the boundary value of 
the dilaton $e^{\f_0}\propto N^{-1}$ to be parametrically small so 
that the restriction above is pushed towards $t\to 0$.    

On the other hand, we may be interested in the strict $N=\infty$
case for a physical reason. For example, in \cite{exotic2} we
propose that the strict $N=\infty$ case corresponds to the XY-spin
models that constitute canonical models of superfluidity. Then, if
we are interested in a holographic model for superfluidity, we can
take the strict $N\to\infty$ limit. In this case one can extend the safe region up to 
$t\to 0^+$. However, there is an analogous problem where the UV region becomes 
highly curved. 

\section{Discussion}
\lab{discuss}

There are several possible directions one can improve on the results obtained here:
\begin{itemize}

\item{Holographic implications in dual field theories}. 

Hawking-Page transitions in gravity generically
correspond to confinement-deconfinement transitions in the dual
gauge theory \cite{Witten}. Therefore, the backgrounds presented
here are natural candidates for continuous type
confinement-deconfinement transitions. More interestingly, it is
also possible to relate gravity to field theories that describe
continuum limit of certain condensed matter systems such as the
spin models. We refer the reader to our companion paper
\cite{exotic2} for a detailed exploration of this latter
application.

\item{Embedding in string theory}

One of main results is that the continuous transition corresponds
to the limit where the BH horizon marginally traps a curvature
singularity where the Einstein's gravity becomes not trustable. We
avoided this problem by fine-tuning the coefficient of the
subleading term in the potential to be very small. This crude way
serves our purpose to determine the qualitative scaling behavior
of observables in the model, but clearly it is desirable to obtain
better control over the model for holographic applications. In
order to resolve the singularity one may think of embedding the
background in the full string theory. Indeed we showed that the
deep-interior asymptotics of the BH geometry near the transition
is a linear-dilaton geometry in the case when the scalar is a
dilaton, which is {\em an exact solution in the string frame, in
all orders in $\a'$}. This observation should be relevant in
embedding the set-up in the full string theory. However we need
more than this because the full flow geometry is not $\a'$ exact
and there is no obvious way to study this solution in a systematic
expansion in the full string theory, because the typical curvature
scale in the geometry is $\ell/\ell_s \sim 1$. Therefore, a
better strategy is to start with a consistent truncation of a
critical string theory and directly search for continuous critical
phenomena there. A helpful starting point for this investigation
may be the work of \cite{Minwalla}\footnote{We thank Yaron Oz for
pointing out the relevance of this work and fruitful discussions
on this issue.}

We also observe that the asymptotic form of the scalar potential
in (\ref{Vs}) and (\ref{case1}) is sum of exponentials that quite
generically appears in consistent truncations of IIB and IIA
critical string theory. We leave this investigation for future
work.

\item{The case when the scalar is not dilaton}

Considering the scalar field as dilaton has various advantages.
First of all, if one would like to think of the framework as an
approximation to non-critical string theory\footnote{This
approximation is viable only if $\ell_s/\ell$ is small so that
neglecting the higher derivative terms can be an approximation. As
we mentioned several times this is not a tunable parameter in our
set-up but rather should be determined phenomenologically by
matching a corresponding quantity in the dual field theory. For
example in certain models of holographic QCD \cite{GKMN3} it turns
out that this parameter is indeed small, about $1/7$.}, then the
most economical choice is to consider $\f$ as the dilaton field of
NCST. Secondly, for this choice the IR asymptotics become the
linear dilaton geometry that is known as a full solution to NCST.
As we show in \cite{exotic2}, another reason is that, in this case
scaling of certain observables in the dual field theory, such as
the magnetization if the dual theory is a spin-model, has expected
scaling properties as a function of $t$. Yet, it would be
interesting to investigate the cases when $\f$ is just a scalar
field, and work out the implications in a hypothetical dual
theory. It is also interesting to search for continuous
Hawking-Page transitions when there are more than one scalars
present.

\item{The specification of the UV geometry}

In this work, we were mostly interested in scaling of
thermodynamic observables near the transition that corresponds to
the IR limit of the geometry. For the various applications in
holography however, the entire background geometry is important.
In order to be able to apply the usual prescription of AdS/CFT to
compute correlation functions, etc. it is desirable to construct a
solution that has the desired IR asymptotics (\ref{Vs}) and an
asymptotically AdS UV geometry at the same time. This is done in 
section 6. One particularly interesting future direction is to find analogous 
solutions in IIB.

\item{The transition in the case $G_D \ne 0$}

One can ask what happens if we turn on (small) gravitational
interactions but ignore the higher derivative corrections. In this
case the entropy of the BH solution (normalized by $1/N^2$) is
given by,
\be\lab{corent} S_{BH} = \frac{e^{(d-1)A(\f_h)}}{4G_D N^2} +
\delta S_{BH}, \ee
where $\delta S$ is the correction that stems from the determinant
around the saddle which is $\cO(G_D)$ with this normalization. The
thermal radiation has a single contribution of the same order
$S_{TG} = \cO(G_D)$. Therefore, the continuous transition would occur 
at $\f_c$ that is determined by
\be\lab{corent2} \frac{e^{(d-1)A(\f_c)}}{4G_D N^2} = S_{TG}-
\delta S_{BH}. \ee
There are two qualitatively distinct solutions to this equation.
In the first case both sides of this equation may vanish. This is
the same situation that we studied in this paper in the classical
limit. If this is what happens then we conclude that the $G_D$
corrections do not help avoiding the singularity and the
transition still happens at the singular point where the scale
factor vanishes. However, there is a second possibility, that the
RHS is finite, hence the transition is moved away from the
singular point by the $G_D$ corrections. In order to find which
possibility is realized one should compute the spectrum of
fluctuations around the TG and the BH saddles at a temperature
determined by $\f_h$ and see if the difference in the entropy can
be matched to the LHS in (\ref{corent}) at a finite $\f_c$. One
should also check that this $\f_c$ does not lie beyond the point
where the Jeans instability destroys the TG saddle.

\item{Adding other fields} 

One can extend the analysis by including other fields in the action (\ref{action}). 
A particularly interesting case to consider is multiple scalar fields. One can ask 
what conditions should be imposed on the scalar potential $V(\f_1, \f_2,\cdots)$ 
in order to obtain continuous Hawking-Page transitions. 

\end{itemize}

\addcontentsline{toc}{section}{Acknowledgments}
\acknowledgments

\noindent It is a pleasure to thank Alex Buchel, Bernard de Wit,
Elias Kiritsis, Yaron Oz, Marco Panero, Erik Plauschinn, Giuseppe
Policastro, Henk Stoof and Stefan Vandoren for valuable discussions. The
author is supported by the VIDI grant 016.069.313 from the Dutch
Organization for Scientific Research (NWO).

\newpage

\appendix
\renewcommand{\theequation}{\thesection.\arabic{equation}}
\addcontentsline{toc}{section}{Appendices}
\section*{APPENDIX}

\section{The scalar variables and reduction of the Einstein-scalar system}\label{AppA}
The scalar variables of the Einstein-scalar system is defined
\cite{GK}\cite{GKMN2} as, In case of a neutral black-hole in $d+1$
dimensions that is solution to (\ref{action}), one only needs two
such variables, that we call $X$ and $Y$:
\begin{equation}\label{XY1}
  X(\f)\equiv \frac{\g}{d}\frac{\f'}{A'}, \qquad  Y(\f)\equiv
  \frac{1}{d}\frac{g'}{A'}.
\end{equation}
where the function $g$ is defined as $g = \log{f}$ and the
constant $\g$ is given by,
\begin{equation}\label{gam1}
  \g = \sqrt{\frac{d~\xi}{d-1}}.
\end{equation}
These functions satisfy:
\bea\label{Xeq1}
\frac{dX}{d\f} &=& -\g~(1-X^2+Y)\le(1+\frac{1}{2\g}\frac{1}{X}\frac{d\log V}{d\f}\ri),\\
\frac{dY}{d\f} &=& -\g~(1-X^2+Y)\frac{Y}{X }. \label{Yeq1} \eea
This second order system is sufficient to determine all of the
thermodynamic properties (and dissipation) of the gravitational
theory \cite{GKMN2}. This is a reduction of the fifth order
Einstein-scalar system to an equivalent second order system.

It is straightforward to show that these equations combined with
the following three,
\bea\label{Ap}
\frac{dA}{dr} &=& -\frac{1}{\ell}e^{A_0} e^{-\g\int^{\f}_{\f_0} (X(t) - \frac{1}{d}X^{-1}(t))dt},\\
\lab{fp} \frac{d\f}{dr} &=&  -\frac{1}{\ell}e^{A_0} \frac{d}{\g} X(\f) e^{-\g\int^{\f}_{\f_0} (X(t) - \frac{1}{d}X^{-1}(t))dt},\\
\lab{gp} \frac{dg}{dr} &=& -\frac{1}{\ell}e^{A_0}~d~Y(\f)
e^{-\g\int^{\f}_{\f_0} (X(t) - \frac{1}{d}X^{-1}(t))dt}, \eea
solve the original Einstein equations. Here $\f_0$ is a cut-off
near the boundary. One can also express $g$ and $A$ in terms of
the scalar variables directly from the definitions (\ref{XY1}):
\bea
A(\f) &=& A_0 + \frac{\g}{d}\int_{\f_0}^{\f} \frac{1}{X}d\tilde{\f},\lab{Aeq1}\\
g(\f) &=&  \g \int_{\f_0}^{\f} \frac{Y}{X}d\tilde{\f}.\lab{feq1}
\eea
Another useful equation relates the scalar potential to the scalar
variables, that follows from (\ref{Xeq1}):
\be\lab{V1} V(\f) = \frac{d(d-1)}{\ell^2} \le(1+Y-X^2\ri)
e^{-2\g\int_{-\f_0}^{\f} (X(t)-\frac{Y(t)}{2X(t)})dt}. \ee
The precise form of the overall coefficient follows from inserting (\ref{Ap}), (\ref{fp}) and (\ref{gp}) in the Einstein equation 
(\ref{E2}). 
 
The zero T solution (thermal gas) corresponds to the $Y=0$ fixed point of eqs. (\ref{Xeq1}) and  (\ref{Yeq1}):
\be\label{Xeq2}
\frac{dX_0}{d\f} = -\g~(1-X^2)\le(1+\frac{1}{2\g}\frac{d\log V}{d\f}\ri).\\
\ee
The solution to this equation  determines the zero T scalar variable $X_0$. In \cite{GKMN2} it is shown that,
\be\lab{Xlim}
\lim_{\f_h\to\infty} X(\f) = X_0(\f).
\ee
Equation  (\ref{Xeq2}) provides a different representation of the scalar potential:
\be\lab{V2} V(\f) = \frac{d(d-1)}{\ell^2} \le(1- X_0^2\ri)
e^{-2\g\int_{-\f_0}^{\f} X_0(t))dt}. \ee

 \section{Derivation of thermodynamics from the scalar variables}
\lab{Apptherm}

The thermodynamics of the black-hole can directly be determined
from the solution in terms of the scalar variables. The free
energy is given by
\be F(\f_h) = \frac{1}{4G_D} \int^{\infty}_{\f_h}
d\tilde{\f}_h~e^{(d-1)A(\f_h)}~\frac{dT}{d\tilde{\f}_h}.\lab{Feq1}
\ee
Our backgrounds satisfy the 1st law of thermodynamics $S =
-dF/dT$. Equation  (\ref{Feq}) directly follows from integrating this
equation  by noting the equation  (\ref{ent3})
\begin{equation}\label{ent31}
S = 4\pi e^{(d-1)A(r_h)} + \cO(N^{-2}),
\end{equation}
and ignoring the $1/N^2$ corrections. The temperature in the
scalar variable systems is obtained from
\be\lab{Teq1} T(\f_h) =
\frac{\ell}{4\pi(d-1)}~e^{A(\f_h)}~V(\f_h)~e^{\g
\int_{\f_0}^{\f_h} X(\f)~d\f}. \ee
Derivation of this equation  is more involved. It follows from the
generalization to arbitrary dimensions, of the ``useful relation''
equation  (7.38) of \cite{GKMN2}. The computation is analogous to the
appendix H.11 of \cite{GKMN2} with slight changes due to different
multiplicative factors in the equations in app. (\ref{AppA})
above. One finds the generalized ``useful relation'' as:
\be\lab{useful} \frac{S}{T^{d-1}} = \frac{1}{4G_D} \le(\frac{4\pi
(d-1) }{\ell}\ri)^{d-1}
\frac{e^{-\g(d-1)\int_{-\infty}^{\f_h}X}}{V(\f_h)^{d-1}}, \ee
from which (\ref{Teq1}) follows immediately by use of
(\ref{ent31}).

\section{Sub-leading corrections to the geometry near $T_c$}\lab{AppC}

The sub-leading corrections are found by expanding the scalar
variables as
\be\label{asXY1} X(\f) = -\frac{1}{\sqrt{d}}+\delta X(\f), \qquad
Y(\f) = Y_0(\f) + \delta Y(\f), \ee
and inserting in (\ref{Xeq}) and (\ref{Yeq}). Then one obtains a
linear differential equation  for $\delta X$ that is solved by fixing
the integration constant with the  requirement of no singularity
at $\f_h$: One finds,
\be\lab{deltaX} \delta X(\f) = \left\{ \begin{array}{ll}
\frac{\tilde{C}\kappa}{\kappa +\tilde{\g}} e^{-\kappa\f} \frac{1-e^{-(\kappa+\tilde{\g})(\f_h-\f)}}{1-e^{-\tilde{\g}(\f_h-\f)}}, & \mathrm{case\,\, i}, \\
\frac{\tilde{C}\a}{\tilde{\g}} \f^{-\a}
\frac{1-e^{-\tilde{\g}(\f_h-\f)}\le(\f_h/\f\ri)^{-\a-1}}{1-e^{-\tilde{\g}(\f_h-\f)}},
& \mathrm{case\,\, ii},
\end{array}\right.
\ee
where we defined,
\be\lab{defss} \tilde{\g} = \frac{\g(d-1)}{\sqrt{d}}, \qquad
\tilde{C} = C \frac{d-1}{2\sqrt{d}}, \ee
and the coefficients $C$, $\kappa$ and $\a$ are defined in
(\ref{case1}) and (\ref{case2}). The expression for $Y$ follows
from the leading term in (\ref{Yeq}):
\be\label{Y0} Y_0(\f) =
\frac{d-1}{d}\frac{1}{e^{2\frac{(d-1)\g}{d}(\f_h-\f)}-1}. \ee
One can also obtain the differential equation for $\delta Y$ from
the sub-leading term in (\ref{Yeq}) but we shall not need it here.
The correction term in (\ref{Aeq}) follows from the expression for
$\delta X$ above. One obtains,
\be\lab{delA}
\delta A(\f) = \left\{ \begin{array}{ll}
\frac{\tilde{C}}{\kappa+\tilde{\g}}e^{-\kappa\f}+\cdots, & \mathrm{case\,\, i}, \\
-\frac{C}{2}\f^{-\a}+\cdots, & \mathrm{case\,\, ii},
\end{array}\right.
\ee
in the limit $\f_h\to\infty$.

\section{Expressing $T_c$ and $\f_\infty$ in terms of $V_\infty$}\lab{vinftc}

Using equation  (\ref{Teq1}), we express $A_s$ as,
\be\lab{D13}
e^{2A_s(\f_h)} = e^{2A(\f_h) + \frac{4}{d-1}\f_h} = \frac{T^2}{V(\f_h)^2}\le(\frac{4\pi (d-1)}{\ell}\ri)^2e^{\frac{4}{d-1} \f_h -2\gamma \int^{\f_h}_0 X }.
\ee
We consider the large $\f_h$ limit of this equation. The potential $V(\f)$ in this limit, is given by (\ref{Vs}). On the other hand, we can represent the
potential $V(\f)$ (for any $\f$) by  equation  (\ref{V2}). In the limit $\f_h\to\infty$ the function $X$ asymptotes to $-1/\sqrt{d}$ as in (\ref{asXY1}). Thus,
by comparing (\ref{V2}) and (\ref{Vs}) in this limit one finds\footnote{ Using also the fact that the functions $X(\f)$ and $X_0(\f)$ become the same in the limit $\f_h\to\infty$.},
\be\lab{D131}
e^{-2\gamma \int_0^{\f_h} X} = V_\infty \frac{\ell^2}{(d-1)^2}e^{ \sqrt{\frac{\xi}{d-1}} \f_h}.
\ee
In the NCST case the normalization of $\f$ is given by $\xi = 4/(d-1)$. Then, substitution of (\ref{D131}) in (\ref{D13}), use of (\ref{Vs}) and the fact that $T\to T_c$ in this limit one finds,
\be\lab{D132}
\lim_{\f_h\to\infty} e^{2A_s(\f_h)} = \frac{(4\pi T_c)^2}{V_\infty}.
\ee
On the other hand we know that $A_s(\f_h)\to 0$ in this limit. Thus, one finds,
\be\lab{D133}
T_c = \frac{\sqrt{V_\infty}}{4\pi},
\ee
or
\be\lab{D151}
V_\infty = (4\pi T_c)^2.
\ee

To express $\f_\infty$ in terms of $V_\infty$ we use the e.o.m. for $\f$, (\ref{fp}). We first recall that the constant $\f_\infty$ is defined as,
\be\lab{D161}
\lim_{r\to\infty} \f(r) = \f_\infty~r.
\ee
Use of  (\ref{Aeq1}) in (\ref{fp}) gives,
\be\lab{D17}
\frac{d\f}{dr} =  -\frac{d}{\ell\g}e^{A(\f)} X(\f) e^{-\g\int^{\f}_{0} X}.
\ee
Now, using (\ref{D131}), (\ref{asXY1}) and the fact that  $\xi = 4/(d-1)$,in the NCST case, one finds in the limit $\f_h\to\infty$,
\be\lab{D18}
\lim_{\f_h\to\infty} \frac{d\f}{dr}\bigg|_{\f=\f_h} = \f_\infty = \frac{\sqrt{dV_\infty}}{\g(d-1)}.
\ee
Using the definition (\ref{gam1}) and $\xi = 4/(d-1)$, one finally arrives at,
\be\lab{D19}
\f_\infty = \frac{\sqrt{V_\infty}}{2}.
\ee

From this one can also express $A_\infty$ in terms of $V_\infty$. We recall the definition of $A_\infty$,
\be\lab{D20}
\lim_{r\to\infty} A(r) = -A_\infty~r.
\ee
Comparison of (\ref{fs}) and (\ref{D20}) gives,
\be\lab{D152}
A_\infty = \frac{2}{d-1}\f_\infty = \frac{\sqrt{V_\infty}}{d-1}= \frac{4\pi}{d-1}~T_c,
\ee
where the last equation  follows from (\ref{D151}).

\section{Curvature invariants}
\lab{AppRicci}

Einstein's equations that follow from (\ref{action}) read,
\be\lab{Einstein} R_{\m\n} - \half g_{\m\n} R = \xi
\le(\6_{\m}\f\6^{\m}\f - \half g_{\m\n} (\6\f)^2 \ri) + \half
g_{\m\n} V(\f). \ee
Tracing both sides one obtains,
\be\lab{Ricci1} R = \xi (\6\f)^2 - \frac{d+1}{d-1}V(\f). \ee
One can express the right hand side in terms of the scalar
variables of Appendix \ref{AppA}. Substituting (\ref{fp}) and
(\ref{V1}) in (\ref{Ricci1}) one finds,
\be\lab{Ricci2} R = \frac{d}{\ell^2} \le[ (d-1)X^2 - (d+1)
(1+Y-X^2) e^{\g\int^{\f}_0 \frac{Y}{X}} \ri] e^{-2\g\int^{\f}_0
X}. \ee
Both for the TG solution and for the BH solution in the limit
$\f_h\gg 1$ (which corresponds to the phase transition region $T
\approx T_c$) one has, $Y\approx 0$, $X\approx X_0$ for all $\f$
\footnote{Except at strict $\f=\f_h$ where in the BH case $Y$ has
a singularity at this point.}. This is indeed the limit that we
are interested in. Equation (\ref{Ricci2}) becomes,
\be\lab{Ricci3} R \approx \frac{d}{\ell^2} \le[ 2d X_0^2 - (d+1)
\ri] e^{-2\g\int^{\f}_0 X_0}. \ee
The approximation becomes better and better as $\f_h$ gets closer
to infinity in the BH case. In the TG case (\ref{Ricci3}) is
exact. We are interested in how $R$ scales with $\f$ for large
$\f$. The biggest contribution come from the large $\f$ region
where we can use the asymptotic expression for the function $X$,
(\ref{asXY1}): $X_0\to -1/\sqrt{d}$. Substituting this in
(\ref{Ricci3}) and using (\ref{D131}) ($X(\f)$ and $X_0(\f)$
become the same functions in the limit $\f_h\gg 1$), one finally
arrives at,
\be\lab{Ricci4} R \approx -\frac{d V_\infty}{d-1}
e^{\frac{2\sqrt{\xi}}{\sqrt{d-1}} \f}, \qquad as\,\,\, \f\gg 1
\ee
To repeat, in the BH case the approximation is valid in the region
$\f_h\gg 1$. We see that $R$ is proportional to $V$ in the IR
region. The same is true for the dilaton invariant $(\6 \f)^2$.
One can also show that the higher order invariants such as
$R_{\m\n\alpha\beta} R^{\m\n\alpha\beta}$ also scale as (powers)
of $V$ in the IR.

One can calculate the string frame invariants from the Einstein-frame ones by passing 
to the string frame metric,
 \be\lab{st1}
g_{s,\m\n} = g_{\m\n} e^{\frac{4}{d-1} \f}.
\ee
Ricci scalar transforms as, 
\be\lab{st2}
R_s = e^{-\frac{4}{d-1} \f}\le[ R + 4 \frac{d}{d-1}\6_{\m}\6^{\m} \f + 4\frac{d}{d-1}\6_{\m}\f\6_{\m}\f \ri],
\ee
where the indices are raised and lowered by the Einstein-frame metric on the RHS. Let us specify to the 
case of NCST normalization $\xi = 4/(d-1)$ where also the relations (\ref{inftc}) hold.  
Using (\ref{Ricci4}) and (\ref{inftc}) we see that the first and the last terms in the square brackets cancel 
in the leading order, wheras the second term scales exactly as the subleading pieces in the first and the last terms. Therefore one finds, 
\be\lab{st3}
R_s \propto \6^2_{r} \delta \f(r),
\ee
where $\delta \f$ is the subleading piece in the IR: $\f \to \f_\infty r + \delta\f(r)$. In order to find this, 
one works out equation (\ref{D17}) to sub-leading order by substituting $X$  in (\ref{asXY1}) and using 
(\ref{deltaX}) in the limit $\f_h\to\infty$. One then finds, 
\be\lab{st4}
\delta \f \propto e^{-\kappa \f_\infty r} 
\ee
in case i and,
\be\lab{st5}
\delta \f \propto r^{-\a} 
\ee
in case ii. Substituion in (\ref{st3}) finally yields, 
\be\lab{st6}
R_s \propto  e^{-\kappa\f} ,\,\,\,\, (case\,\,i); \qquad \f^{-\a} ,\,\,\,\, (case\,\,ii),
\ee
in the limit $\f\to\infty$. We see that the string frame Ricci scalar vanishes near the transition region. 
The same is true for the dilaton invariant  $(\6 \f)^2$ in the string frame. Higher order invariants 
vanish with a stronger rate.  This shows that the $\a'$ corrections in the string frame become unimportant near the transition region.

\section{Bulk fluctuations and dissipation}
\lab{AppF}

In this section, we present the details of the calculations in section \ref{bulkdeform}. We first derive the fluctuation equation  (\ref{fluceq1}) that we reproduce here:
\be\lab{F1}
h_b'' +h_b' \le( \frac{2\xi}{d-1}\frac{1}{A'} + (d-2)A' - 3B' + \frac{f'}{f}\ri) + h_b\le( \frac{\o^2}{f^2} e^{2B-2A} + \frac{f'}{f}B' - \frac{\xi}{d-1}\frac{f'}{f A'}\ri)=0,
\ee
where prime denotes the derivative w.r.t. $\f$ and the metric functions are defined in the $\f$-frame:
\be\lab{F2}
ds^2 = e^{2A(\f)} \le(-f(\f) dt^2 + dx_{d-1}^2\ri) + e^{2B(\f)} \frac{d\f^2}{f(\f)}.
\ee
One fluctuates the Einstein's e.o.m. with the metric fluctuations:
\bea
g_{00} &=& - e^{2A} f \le(1 + \frac{\eps}{2} h_{00} e^{-i\o t} \ri)^2,\nn\\
g_{ii} &=&  e^{2A}  \le(1 + \frac{\eps}{2} h_b e^{-i\o t} \ri)^2,\lab{F3}\\
g_{\f\f} &=&  \frac{e^{2B}}{f}  \le(1 + \frac{\eps}{2} h_{\f\f} e^{-i\o t} \ri)^2,\nn
\eea
and $\delta \f = 0$. The last choice corresponds to the gauge (\ref{F2}).
We need only two of the zeroth order Einstein eqs.  in this gauge:
\bea
f'' + d A'f' - B'f' &=& 0 \lab{F4}\\
A'' - A'B' + \frac{\xi}{d-1} &=&  0.\lab{F5}
\eea
The fluctuation equations for  $h_{00}$, $h_b$ and  $h_{\f\f}$ are obtained by expanding the Einstein eqs.
$G_{\m}^{\n} = T_{\m}^{\n}$ to order $\eps$ where the stress tensor at this order reads:
\be\lab{F6}
T_{\m}^{\n}\bigg|_{\eps} = - \xi e^{-2B} f h_{\f\f} e^{-i\o t} \le( \delta_{\f}^\n\delta^{\f}_\m -\half\delta_{\m}^\n \ri).
\ee
One also needs to expand the Einstein tensors to this order. This is done with help of Mathematica. We shall present the
final results.
From the first order terms in $\eps$ in the Einstein's
equations one finds the following e.o.m for the fluctuations: From
the $\{\f,\f\}$ Einstein equation  one finds,
\be\lab{F10} h_{00}' = h_{\f\f} \le(-\frac{\xi}{d-1}\frac{1}{A'} +
d A'+ \frac{f}{f'}\ri) -
h_b'\le(d-1+\frac{f'}{2A'f}\ri)-\frac{\o^2}{f^2}\frac{e^{2B-2A}}{A'}h_b.
\ee
 From the $\{0,\f\}$ Einstein equation  one finds,
\be\lab{F11} h_{\f\f} = \frac{1}{A'}\le(h_b'-\frac{f'}{2f}
h_b\ri). \ee
Finally, from the $\{i,i\}$ Einstein equation  one finds a complicated
expression that involves $h_b$, $h_b'$ and $h_b''$ that simplifies
upon use of (\ref{F10}), (\ref{F11}) and the zeroth order eqs.
(\ref{F4}) and (\ref{F5}) and reduces to (\ref{F1}).

Now, we move on to the details that lead to (\ref{hbflux}) that we
reproduce here:
\be\lab{F12} Im~J =
\frac{i}{4}\xi\frac{f}{A^{'2}}e^{dA-B}\le(h_b^*h_b'-h_b
h_b^{*'}\ri). \ee
This is obtained by expanding the Lagrangian (\ref{action}) to
second order in $\eps$ in fluctuations, as explained in
\cite{GubserBulk}, as follows:

Define the fluctuation vectors $\vh = (h_{00},h_b,h_{\f\f})$. Then
the $\cO(\eps^2)$ term in the Lagrangian is of the form,
\be\lab{F13} {\cal L} = \half \dot{h}^T~M_{tt}~h +  \half
h^{'T}~M_{\f\f}~h'  + h^{'T}~M_{\f}~h  + \half h^{T}~M~h + \cdots
\ee
where the ellipsis refer to total derivative terms in $\f$ and in
t. Now, one formally promotes $\vh$ to {\em complex valued} fields
and writes ($h$ should be understood as a complex vector in the
following):
\be\lab{F14} {\cal L} =   h^{\dag '}~M_{\f\f}~h' + h^{\dag
'}~M_{\f}~h + \cdots \ee
where remaining terms are unimportant as they do not contribute to
$Im~J$ which is the conserved flux under the $U(1)$ rotations
$\vh\to e^{i\theta} \vh$:
\be\lab{F15} {\cal L} =   \6_\f J + \cdots, \qquad J = h^{\dag}
\le( M_{\f\f}~h' + (M_{\f}+ \cdots)~h\ri). \ee
Here the ellipsis in ${\cal L}$ refer to terms that vanish on
shell and the ellipsis in $J$ refer to a symmetric matrix that
vanish in the imaginary part of $J$ hence irrelevant for our
purpose.

Thus the entire calculation is reduced to the computation of the
$3\times 3$ matrices $M_{\f\f}$ and $M_{\f}$. Clearly, only the
$\sqrt{g}R$ piece of (\ref{action}) will give rise to terms with
single or double derivatives in $\f$, as in (\ref{F13}). Moreover,
we are not interested in the t-dependence, therefore one can
ignore the time dependence in the fluctuations (\ref{F3}). One
expands $\sqrt{g}R$ with help of a symbolic computer code and
obtains $M_{\f\f}$ and $M_{\f}$ as (after integration by parts in
the double derivative terms and upon use of the zeroth order eqs.
(\ref{F4}) and (\ref{F5})):
\begin{displaymath}
M_{\f\f} = (d-1)~f~e^{dA-B}\left( \begin{array}{ccc}
0 & \half & 0 \\
\half & \frac{d-2}{2} & 0 \\
0 & 0 & 0
\end{array} \right), \quad
M_{\f} = \frac{(d-1)}{4}~e^{dA-B}\left( \begin{array}{ccc}
* & -f'-2A'f & \frac{f'+2A'f}{d-1} \\
-2fA' & * & 2fA' \\
\frac{f'+2dA'f}{d-1} & f'+2dA'f & *
\end{array} \right)
\end{displaymath}
where the order of the components are according to $\vh =
(h_{00},h_b,h_{\f\f})$ and the stars refer to terms that do not
contribute to the imaginary part of the flux. One calculates the
flux from (\ref{F15}). It reduces to (\ref{F12}) upon use of the
first order equations (\ref{F10}) and (\ref{F11}).


\end{document}